\documentclass[traditabstract]{aa} 
\usepackage{graphicx}
\usepackage{txfonts}
\usepackage{natbib}
\usepackage{lscape} 
\bibpunct{(}{)}{;}{a}{}{,} 
\begin{document}
%
   \title{The VIMOS Ultra-Deep Survey: evidence for AGN feedback 
in galaxies with CIII]-$\lambda$1908\AA ~emission 10.8 to 12.5 Gyr ago
	  \thanks{Based on data obtained with the European 
	  Southern Observatory Very Large Telescope, Paranal, Chile, under Large
	  Program 185.A--0791. }
}

\author{O.~Le F\`evre\inst{1}
\and B.~C.~Lemaux \inst{1,2}
\and K. Nakajima\inst{3,4}
\and D. Schaerer\inst{3,5}
\and M. Talia\inst{6}
\and G. Zamorani\inst{7}
\and P.~Cassata\inst{5}
\and B.~Garilli\inst{8}
\and D.~Maccagni\inst{8}
\and L.~Pentericci\inst{9}
\and L. A. M.~Tasca\inst{1}
\and E.~Zucca\inst{7}
\and R.~Amorin\inst{9,13}
\and S.~Bardelli\inst{7}
\and A.~Cimatti\inst{6}
\and M.~Giavalisco\inst{10}
\and L.~Guaita\inst{9}
\and N.P.~Hathi\inst{11}
\and F.~Marchi\inst{9}
\and E.~Vanzella\inst{7}
\and D.~Vergani\inst{9}
\and J.~Dunlop\inst{12}
}

\institute{Aix Marseille Universit\'e, CNRS, LAM (Laboratoire d'Astrophysique de Marseille) UMR 7326, 13388, Marseille, France
\and
Department  of  Physics,  University  of  California,  Davis,  One  Shields  Ave.,  Davis,  CA  95616,  USA
\and
Geneva Observatory, University of Geneva, ch. des Maillettes 51, CH-1290 Versoix, Switzerland
\and
European Southern Observatory, G-85748 Garching bi Munchen, Germany 
\and
Institut de Recherche en Astrophysique et Plan\'etologie - IRAP, CNRS, Universit\'e de Toulouse, UPS-OMP, 14, avenue E. Belin, F31400
Toulouse, France
\and
University of Bologna, Department of Physics and Astronomy (DIFA), V.le Berti Pichat, 6/2 - 40127, Bologna
\and
INAF--Osservatorio Astronomico di Bologna, via Ranzani, 1 - 40127, Bologna
\and
INAF--IASF Milano, via Bassini 15, I--20133, Milano, Italy
\and
INAF--Osservatorio Astronomico di Roma, via di Frascati 33, I-00040, Monte Porzio Catone, Italy
\and
Astronomy Department, University of Massachusetts, Amherst, MA 01003, USA
\and
Space Telescope Science Institute, 3700 San Martin Drive, Baltimore, MD 21218, USA
\and
SUPA, Institute for Astronomy, University of Edinburgh, Royal Observatory, Edinburgh, EH9 3HJ, United Kingdom
\and
Kavli Institute for Cosmology, University of Cambridge, Madingley Road, Cambridge, CB3
 \\ \\
             \email{olivier.lefevre@lam.fr}
}

   \date{Received ...; accepted ...} 

%
 
  \abstract{
We analyze the CIII]-$\lambda$1908\AA ~emission properties in a sample of 3899  star-forming galaxies (SFGs) at 2$<$$z$$<$3.8  drawn from the VIMOS Ultra-Deep Survey (VUDS). 
We find a median equivalent width $EW(CIII])$=2.0$\pm$0.2 to 2.2$\pm$0.2\AA ~for the whole SFG population at 2$<$$z$$<$3 and 3$<$$z$$<$4, respectively. About 24\% of SFGs are showing $EW(CIII])$$>$3\AA, 
including $\sim$20\% with modest emission  3$<$$EW(CIII])$$<$10\AA ~and $\sim$4\% with strong emission $EW(CIII])$$>$10\AA.
A small but significant fraction of 1.2\% of SFGs presents strong CIII] emission 20$<$$EW(CIII])$$<$40\AA, with the four strongest emitters (EW(CIII])$>$40\AA ~up to $\sim$95\AA) being associated with   broad-line quasars.
While this makes CIII] the second most-frequent emission line in the UV rest-frame spectra of  SFGs after Lyman-$\alpha$, this line alone cannot be considered an efficient substitute to measure a galaxy redshift in the absence of Ly$\alpha$ emission, unless the spectral resolution is R$>$3000 to distinguish among different possible doublets. 
We find a large dispersion in the weak correlation between $EW(CIII])$ and $EW(Ly\alpha)$, with galaxies showing strong CIII] and no Ly$\alpha$, and vice-versa.  
The spectra of SFGs with 10$<$EW(CIII])$<$40\AA ~present strong emission lines including CIV-$\lambda$1549, HeII-$\lambda$1640, OIII-$\lambda$1664, but also weaker emission features of highly ionized elements like SiIV-$\lambda$1403, NIV-$\lambda$1485, NIII-$\lambda$1750, or SiIII-$\lambda$1888, indicating the presence of a hard radiation field. 
We present a broad range of observational evidence supporting the presence of AGN in the strong CIII] emitting population. As EW(CIII]) is rising, we identify increasingly powerful outflows with velocities up to $\sim$1014 km/s, beyond what stellar winds are commonly producing. The strongest CIII] emitters are preferentially located below the main sequence of star-forming galaxies, 
with the median star formation rate being reduced by  a factor of two. In addition, we find that the median stellar age of the strongest emitters is $\sim$0.8 Gyr, about three times that of galaxies with  $EW(CIII])$$<$10\AA.
X-ray stacked imaging of the strong CIII] emitters sample show a marginal 2$\sigma$ detection consistent with low luminosity AGN  $log(L_X(2-10keV))$$\sim$42.9 erg/s. Spectral line analysis and classification presented in a joint paper by Nakajima et al. (2017) confirms that the strongest emitters require the presence of an AGN. We conclude that these properties are indicative of AGN feedback acting in  SFGs at 2$<$z$<$3.8, contributing to star-formation quenching. We find that quenching timescales of $\sim0.25-0.5\times10^9$ years, 
are necessary for such AGN feedback to turn part of the star-forming galaxy population with M$_{star}>10^{10}$M$_{\odot}$ at z$\sim$3 into the  population of  quiescent galaxies observed at redshift $z\sim1-2$.  
}


\authorrunning{Olivier Le F\`evre et al.}
\titlerunning{VUDS: CIII]-$\lambda$1908\AA ~emitters at $2<z<3.8$ and evidence for AGN quenching star-formation }

   \maketitle


\section{Introduction}

Progress in understanding the formation and early assembly of galaxies requires assembling large samples of high redshift galaxies. The first galaxies formed during the epoch of reionization (EoR) are thought to contribute most of the photons transforming neutral Hydrogen into a fully ionized medium at $z>6$ \citep{Robertson:2015}. However,  this evidence is based mostly on a census of galaxies identified from their photometric properties \citep[e.g.][]{Bouwens:14},  while confirming their redshift  remains difficult  because spectral signatures allowing their identification are scarce \citep{Pentericci:2011,Stark2010}. Transforming the large population of galaxy candidates at $z>6$ into reference samples with the confirmed spectroscopic redshifts needed to infer their physical properties remains a major issue.

In the rest-frame UV spectra of galaxies observed at these redshifts, several spectral features are expected to be present, but only a few  emission lines may reach a sufficient strength to be detected and either serve as a useful reference for a robust redshift measurement, or as a physical diagnostic to the ionizing field and interstellar medium (ISM). The properties of these lines need to be well understood to make progress in finding and characterizing galaxies at these redshifts. 

The main spectral features used to identify distant galaxies rely on the properties of the Hydrogen atom between the Lyman$-\alpha$ (hereafter Ly$\alpha$) transition at $\lambda$1215.7\AA ~and the Lyman limit at $\lambda$912\AA. 
While the Lyman-limit and continuum drop-out from  absorption by gas clouds in the inter-galactic medium (IGM) on the line of sight allow the identification of candidate high redshift galaxies from photometry alone, the key feature allowing to robustly identify distant galaxies from spectroscopic follow-up is the Ly$\alpha$ emission, primarily produced by  gas photoionized by young hot stars. 
Ly$\alpha$ emission, when present, is easier to detect than a continuum break and therefore most of the galaxies at z$>$6 securely confirmed from spectroscopy today are based on the detection of Ly$\alpha$ in emission. Narrow-band imaging is also used for  Ly$\alpha$ emitters searches \citep[e.g.][]{Ouchi:2008}, but it remains difficult to analyze extreme populations like strong emitters appearing as massive, old, and with low star formation \citep{Taniguchi:2015} without additional spectral information besides Ly$\alpha$.

Unfortunately, it appears that the fraction of Ly$\alpha$ emitters, while rising up to $z\sim6$ \citep[e.g.][]{Stark:09,Cassata:15,debarros:17}, seems to be suddenly decreasing at higher redshifts, which is interpreted as the signature of the end of reionization at around $z\sim6$ \citep{Stark2010,Pentericci2014}. Neutral Hydrogen acts to significantly scatter Ly$\alpha$ which diffuses the emission, and, in the presence of dust,causes attenuation which makes the feature difficult to detect at $z>6$ \citep{Vanzella2011,Treu2013,Pentericci2014,Oesch:2016,Hoag2017}. 
One is therefore seeking alternatives to Ly$\alpha$ to obtain accurate redshift measurements of the most distant galaxies. 

Interestingly, the Carbon line emission may come to the rescue, from atomic transitions visible both in the sub-millimeter domain with the [CII]-158$\mu$m line \citep[e.g.][]{Carilli2013,Capak2015,Pentericci2016,Bradac2017}, as well as in the UV rest-frame domain with the CIII] emission at $\lambda$$\sim$1908\AA, combined with the more complex CIV-$\lambda$1549 feature, observed both in emission from photo-ionization processes and in absorption \citep{Schmidt2017}. In the UV rest-frame the HeII-$\lambda$1640\AA ~line is also expected to be present as a tracer of the first stellar populations like PopIII present only on short timescales of a few million years \citep[e.g.][]{Schaerer2003,Cassata2013,Sobral2015}, or from young stellar populations \citep{Cassata2013,Kehrig:2015}. 

The CIII]-$\lambda$1908 emission was recently proposed as an alternative to Ly$\alpha$ to identify distant galaxies for which there is no Ly$\alpha$ in emission \citep{Stark2014}. At intermediate redshifts, CIII] emission has been reported starting with the first laarge spectroscopic census of LBGs \citep{Shapley2003}. At these redshifts, this line seems to be more prominent for low-mass systems, and may be correlated to Ly$\alpha$ \citep{Stark2014}. Several studies  report CIII] emission at $2<z<4$, either from individual galaxies  \citep{Erb2010,Talia2012,Karman2014,Steidel2014}, or in high signal-to-noise stacks \citep{Shapley2003}, independently of  whether Ly$\alpha$ is in emission or not \citep{LeFevre:15}. At $z>6$ CIII] is now reported in the observed spectra of several galaxies \citep{Stark2015}, but absent in others \citep{Sobral2015,Schmidt2016} with solid upper limits \citep{shibuya17}. However, the prevalence of CIII] emitters is as yet unknown and a robust reference to the properties of low to intermediate redshift analogs to the high redshift CIII] emitters is therefore needed for this line to be useful in analyzing star-forming galaxies at the highest redshifts \citep{Rigby2015,Du2016,Stroe2017,Amorin2017}.

The CIII]-$\lambda$1908 emission offers an interesting opportunity for a diagnostic of the physical conditions in distant star-forming galaxies. It is a doublet, combination of [CIII]-$\lambda$1907\AA, a forbidden magnetic quadrupole transition and CIII]-$\lambda$1909, a semi-forbidden electro-dipole transition \citep{Stark2014}, with a critical density $N_{e}$$\approx$$10^{10}$cm$^{-3}$ and requiring an ionizing potential of 24.4 eV. In star-forming galaxies the strong ionizing flux produced by young hot stars may be sufficient to photo-ionize Carbon with the production of CIII] photons. The strong ionizing background produced by active galactic nuclei (AGN) also leads to the ubiquitous CIII] emission observed in quasar spectra \citep{VandenBerk2001}. Strong CIII] emission with an equivalent width EW(CIII]) in excess of 20\AA ~can mainly be produced by the hard ionizing spectrum of an AGN, or, in exceptional cases, by extreme stellar populations, as described in \citet{Nakajima:2017}.   
Sufficient ionizing flux may also arise in the presence of shocks, e.g. produced by SNe, jets from a radio-source \citep[e.g. ][]{Best2000}, or from gas compression during a merger event \citep{Bournaud:11}. 
Characterizing the source of CIII] emission can therefore provide important clues on the star formation and the co-evolution with AGN, and in particular the possibility of AGN activity quenching star formation, a mechanism proposed to evolve star-forming galaxies at high-z into quiescent galaxies at $z\sim1-2$ and below \citep{Silk1998} \citep[for a recent account see e.g.][]{Barro2013,Dubois2013}. 

In this paper we seek to characterize the prevalence and properties of CIII]-$\lambda$1908 ~emitters in the general star-forming galaxy population at redshifts $2<z<3.8$ using rest-frame UV spectra obtained from the large VIMOS Ultra Deep Survey  \citep[VUDS, ][]{LeFevre:15}. The large sample of 3899 galaxies with the most reliable redshift measurements in VUDS,  representative of the star-forming galaxy population at these redshifts, enables for the first time to establish the distribution and properties of the CIII] emission 10.8 to 12.5 Gyr back in cosmic time. 

We summarize the VUDS survey properties relevant for CIII] observations in Sect.\ref{survey}, and present the methodology to measure CIII] equivalent width and flux together with examples of individual CIII] emitters of different strengths. The average spectral properties of the different CIII] populations are described in Sect.\ref{avg_c3}, including an attempt to identify the contribution from AGN. The Ly$-\alpha$ properties of CIII] emitters are discussed in Sect.\ref{Lya_c3}. We derive the equivalent width distribution of CIII] emitters and the CIII] emitters fraction among young star-forming galaxies in Sect.\ref{distrib}.  We discuss the general observed properties of the CIII] emitters  in Sect.\ref{properties} including their position on the SFR-M$_{star}$ plane.
Evidence for AGN feedback quenching star-formation  is presented in Sect.\ref{agn_quench}. 
A summary is provided in Sect.\ref{summary}.

The data presented in this paper are further analyzed in light of a wide grid of photoionization models, as discussed by \citet{Nakajima:2017}.

When quoting absolute quantities we use a cosmology with $H_0=70~km~s^{-1}~Mpc^{-1}$, $\Omega_{0,\Lambda}=0.7$ and $\Omega_{0,m}=0.3$. 
All magnitudes are given in the AB system. We use a definition of equivalent width with positive values indicating emission and negative values indicating absorption. 
All equivalent width measurements are given in the rest frame. 


\section{CIII]-$\lambda$1908\AA ~emission in VUDS}
\label{survey}

\subsection{Detecting CIII] in the VUDS spectra}

Spectra of $\approx$10\,000 objects have been obtained for  VUDS to study galaxy evolution in the redshift range $2<z<6+$.  The reader is referred to  \citet{LeFevre:15} for a detailed description of the survey observations, and methods applied to process the data and measure key parameters, including the spectroscopic redshift $z_{spec}$, as well as \cite{Tasca16} for a description of the VUDS-DR1 first data release.

The key element of this sample is that the spectroscopic targets are mainly selected based on their photometric redshifts which satisfy $z_{phot}+1\sigma \geq 2.4$ and $i_{AB} \leq 25$, with a well-defined selection of objects. The wavelength range of the final spectra is $3600 < \lambda < 9350$\AA, cumulating 14h integrations in each of the LRBLUE and LRRED grisms of the VIMOS spectrograph on the ESO Very Large Telescope \citep{LeFevre:03}, with a spectral resolution $R\sim230$. 

Standard data processing is performed using the VIPGI environment \citep{Scodeggio:05}, followed by redshift measurement using the EZ package \citep{Garilli:10}. The target selection produces a final VUDS sample representative of the star-forming galaxy population in the redshift range $2<z<6.5$, flux--selected from the rest-frame continuum UV luminosity.
An important aspect of the VUDS is the wide field (large volume) covered. A total of 1 deg$^2$ is covered in three fields including the COSMOS, ECDFS and VVDS-02h fields. This enables to reduce cosmic variance
in the redshift range of interest, to below 10\% \citep{moster2011}. We identify galaxies with Ly$\alpha$ in emission as well as in absorption \citep[e.g. Figure 19 and 20]{LeFevre:15}, and the fraction of Ly$\alpha$ emitters with EW(Ly$\alpha$)$>$25\AA ~in the redshift range of this study is $\sim11$\% \citep{Cassata:15}.

The instrumental setup translates to the ability to follow the CIII] line reliably from a redshift $z=2$ up to $z\simeq3.8$. At the spectral resolution of the observations, the CIII] $\lambda$1907--1909 doublet is not resolved, and for the remainder of this paper, we will refer to the doublet simply as CIII]-$\lambda$1908, or CIII]. The ability to identify CIII] varies with the observed wavelength in a non-linear way, as the line will appear on top or in-between the strong night-sky OH emission. This aspect is further discussed below. All other lines bluer than CIII] and redder than Ly$\alpha$ useful for spectral diagnostics can be followed for $2<z<3.8$. We are therefore restricting the VUDS sample to the 3899 galaxies observed in the redshift range $2<z<3.8$, with spectroscopic redshift measurements going from reliable (flag 2 and 9, $\sim$80\% probability of being right) to very reliable \citep[flag 3, 4, $\sim$100\% probability of being right, ][]{LeFevre:15}. 

A large range of ancillary data is available for galaxies in VUDS, with physical parameters obtained from spectral energy distribution (SED) fitting of the multi-wavelength photometric data \citep{Tasca2015,Thomas2016},  including stellar mass, star formation rate, dust extinction E(B-V), age, or metallicity. The possible presence of obscured AGN in the sample, with an obscured view of the central engine,  is not expected to strongly contribute to the overall UV continuum \citep[e.g.][]{Bundy2008,Assef2010,Hainline2012}. The physical properties of galaxies derived from SED-fitting using templates which do or do not include an obscured AGN component are compared in \citet{Bongiorno2012} who find little differences on stellar masses and star formation rates.  
Galaxy sizes used in this study are drawn from those measurements reported in \cite{Ribeiro2016} and we also draw on the morphology analysis of bright clumps made in \cite{Ribeiro2017}.


\subsection{Measurements of CIII]-$\lambda$1908 ~equivalent width and flux}
\label{c3_measurement}

The measurement of line flux and equivalent width (EW) is performed on each galaxy individually. We first measure the flux and EW of CIII] using a custom code in IDL to define the sample of CIII] emitters \citep[following][]{Lemaux2010}. When CIII] falls in wavelength domains with moderate to strong OH sky lines,  we flag galaxies for which this is the case in the following way: we raise a 'sky flag' when any such OH sky line \citep[as given in][]{Hanuschik2003} falls within a 17\AA ~bandpass defining the CIII] line (we use bandpasses of 25\AA ~and 23\AA ~for HeII and CIV, resp.). We then iterate manually on these measurements performing a direct line integration  using the {\texttt splot} tool in {\sc IRAF} to obtain the final measurements of EW and flux, after the identification of the continuum under CIII] evaluating the continuum level on each side of the line, and updating the automated measurement if necessary.  We proceed in the same way to obtain measurements of CIV-$\lambda$1549, HeII-$\lambda$1640 ~and other lines of interest, while for Ly$\alpha$ we use a similar technique but perform the continuum estimate from the red side of the line only.  

We estimate the median error in measuring the EW of a line as the r.m.s. uncertainty resulting from noise on the spectra as measured on each side of the line, and we typically measure 1$\sigma_{median}(EW)=1.0$\AA ~for all lines of interest. This results into a $3\sigma$ lower limit of 3\AA ~on EW(CIII]),  limited by the faintness of the sources as observed in low $R\simeq230$ resolution spectra. 



For 2543 galaxies the background noise at the wavelength of CIII] is not affected by sky residuals or other instrumental defects. We find that 1763 of these galaxies have CIII] in emission, EW(CIII])$>0$,
while 429 and 120 galaxies have a signal-to-noise of the CIII] line larger than 3 and 5, respectively.


%

\section{Average CIII] properties from stacked VUDS spectra}
\label{avg_c3}

\subsection{CIII] populations}
\label{c3_pop}

To analyse the CIII] emitters, we arbitrarily split the VUDS sample of star-forming galaxies with $2<z<3.8$ and S/N$>$3 in the following sub-samples:
\begin{itemize}
\item $EW(CIII])>$20\AA: the "very strong emitters"  (31 galaxies)
\item $10<EW(CIII])\leq20$\AA: the "strong emitters" (109 galaxies)
\item $5<EW(CIII])\leq10$\AA: the "medium emitters" (289 galaxies)
\item $0<EW(CIII])\leq$5\AA: the "weak emitters" (1334 galaxies)
\item "SFG": A UV-rest selected representative sample of 1455 star forming galaxies from VUDS (with the best reliability flags 3 and 4), including 951 with $2<z<3$ and 504 with $3<z<4$
\end{itemize}

The median EW(CIII]) for the first four categories are 27.1, 13.8, 8.0, and 3.4\AA, respectively. This compares to EW(CIII])$=2.0\pm0.2$ and $2.2\pm0.2$\AA ~for the whole population of UV-rest selected SFGs at $2<z<3$ and $3<z<4$, respectively.

\begin{figure*}
   \centering
   \includegraphics[width=14cm]{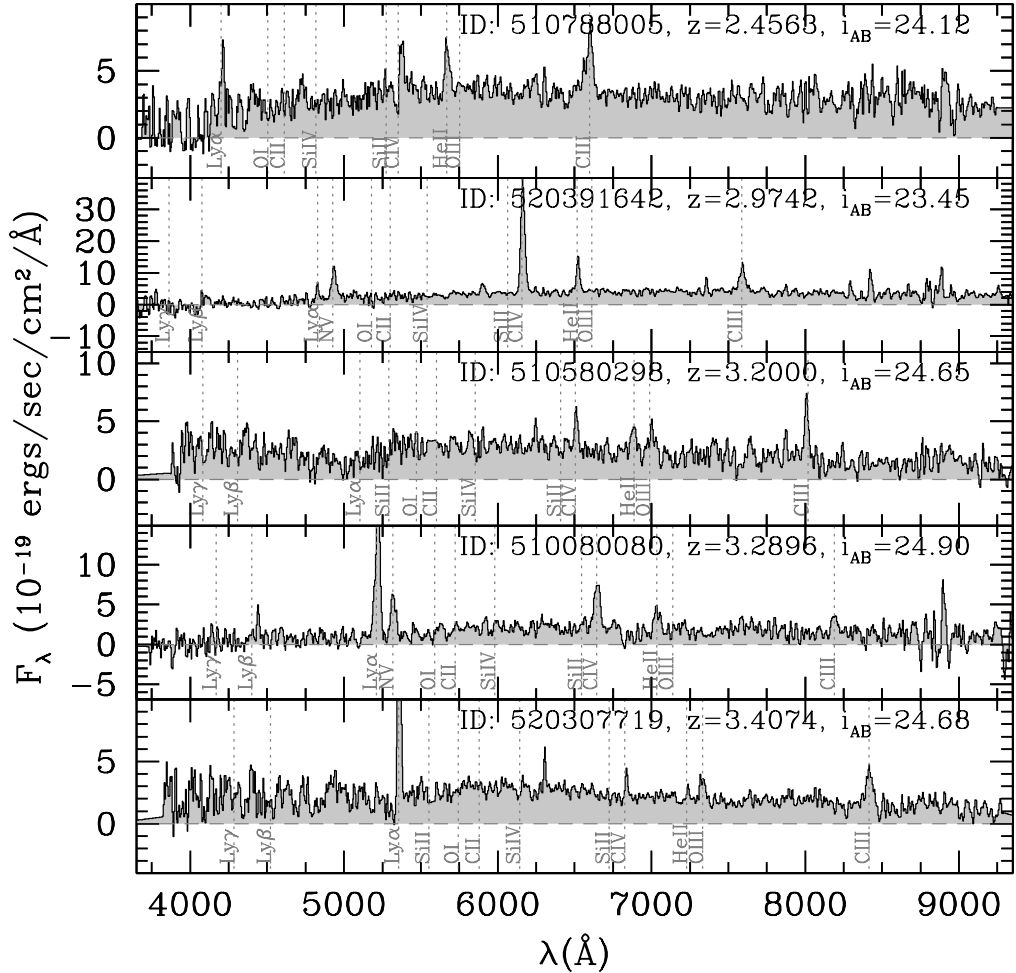}
      \caption{Example VUDS spectra of CIII]1908 emitters with very strong EW(CIII]) $\geq$20\AA. Note the diversity of strong CIII] emitters spectra, including strong Ly$-\alpha$ emission as well as weak Ly$-\alpha$ emission (ID:510788005 and ID:510580298).  
}  
         \label{strong_c3}
   \end{figure*}

\begin{figure*}
   \centering
   \includegraphics[width=14cm]{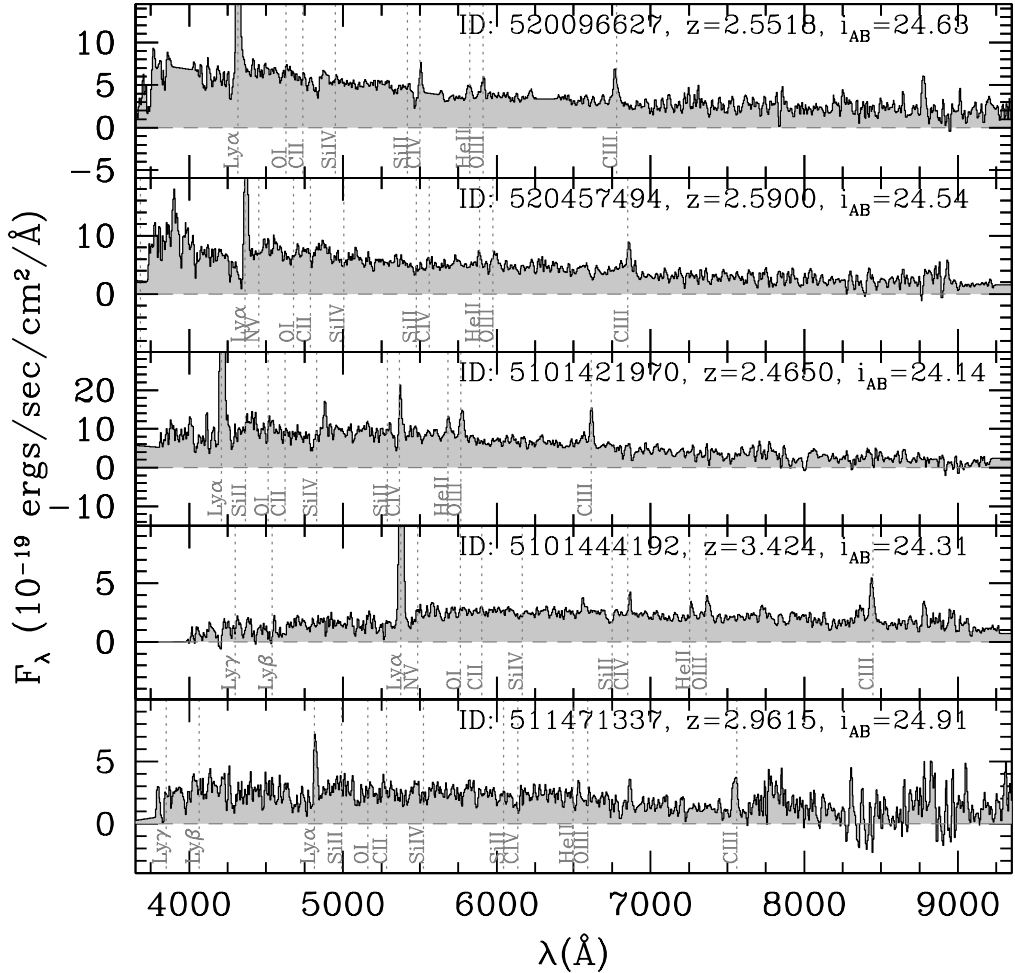}
      \caption{Example VUDS spectra of strong CIII]1908 emitters with 10$<$EW(CIII])$<$20\AA.  
}  
         \label{inter_c3}
   \end{figure*}

\begin{figure*}
   \centering
   \includegraphics[width=14cm]{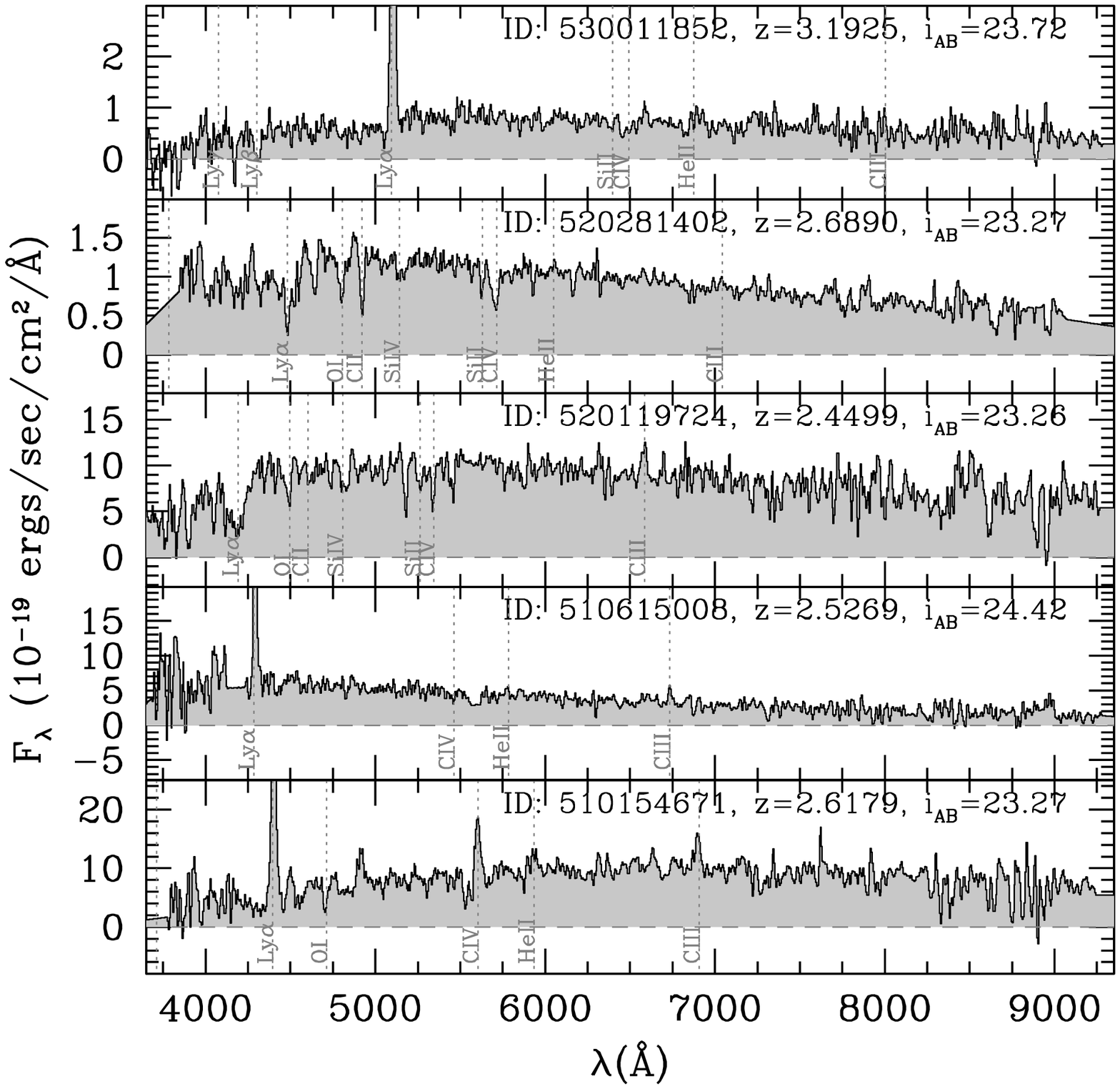}
      \caption{Example VUDS spectra of weak CIII]1908 emitters with 0$<$EW(CIII])$<$5\AA.  
}  
         \label{weak_c3}
   \end{figure*}

Examples of  CIII] emitters over the range of observed EW(CIII]) are presented in Figs. \ref{strong_c3} to \ref{weak_c3}.  

Individual spectra of CIII] emitters show a wide range of properties. 
We observe that CIII] emission occurs in galaxies with Ly$\alpha$ either in emission or absorption. This is further investigated in Sect. \ref{Lya_c3}. 
The presence of CIV-$\lambda$1549 emission indicates a substantial flux of ionizing photons with energies in excess of 47.9 eV. 
Depending on the spectra, and best identified in the stacks (see below), a wealth of absorption and emission lines are identified. The most common emission lines beside CIII] and Ly$\alpha$ are CIV-$\lambda$1549 (a blend of CIV-1549 and CIV-1551, combining nebular emission and stellar absorption) and HeII-1640, and we also observe OIIII]-$\lambda$1663 (blend of the $\lambda$1661-1666 doublet at the resolution of VUDS), NIII-$\lambda$1750, NIV-$\lambda$1485, as well as NV-$\lambda$1240,  SiII-$\lambda$1309, SiIII-$\lambda$1888 and SiIV-$\lambda$1403. In rare cases Ly$\beta$ and OVI-$\lambda$1035 are also observed in emission.

Similarly to \cite{LeFevre:15} we produce summed ("stacked") spectra of VUDS galaxies in the rest-frame using the VUDS catalog redshift \citep[as described in the presentation of the first VUDS data release,][]{Tasca16}. These stacks are produced by normalizing the continuum between 1300 and 1600\AA ~and summing the flux from all galaxies in a given category defined above, using the \texttt{scombine} task under IRAF, with sigma clipping set at $3\sigma$. In these stacks, we do not use objects identified as AGN, either with broad line or NV-$\lambda$1240 emission presented in Sect. \ref{sect_agn}.

\begin{figure*}
   \centering
   \includegraphics[width=\hsize]{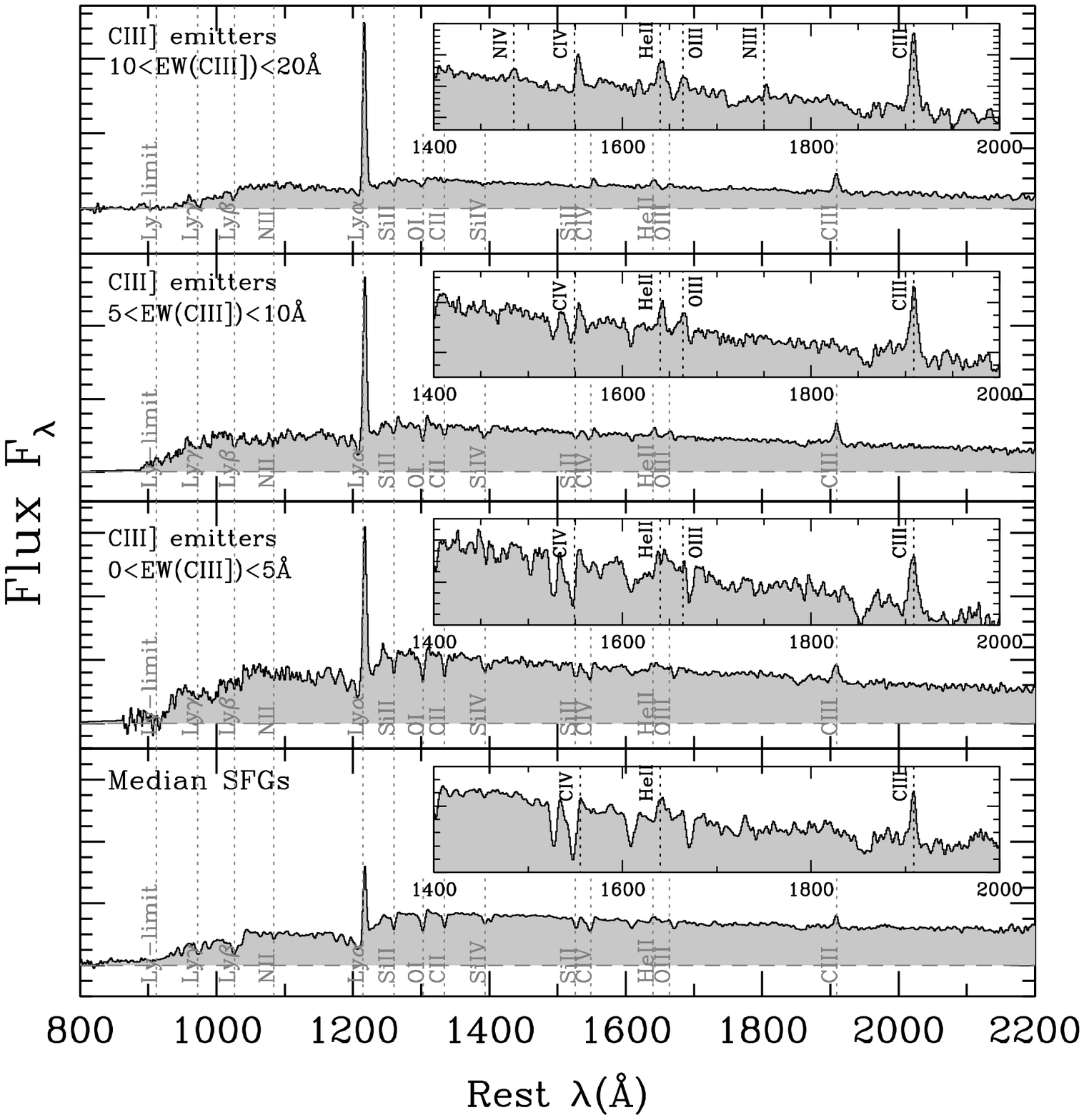}
      \caption{Composite spectra of CIII]1908 emitters identified in the VUDS survey at $2<z<3.8$. From top to bottom: stack of 43 emitters with 10$<$EW(CIII])$<$20, 31 emitters with 5$<$EW(CIII])$<$10, 30 emitters with 0$<$EW(CIII])$<$5, and, for comparison, a stack of 450 galaxies representative of the general SFG population in the redshift range $2.8<z<3.8$.
}  
         \label{stacks_c3}
   \end{figure*}

We find that  the lower EW measurements include significant contamination from noise, that we correct to obtain the fraction of emitters as a function of EW(CIII]) as described in Sect.\ref{distrib}. Spectra for each of the stacks are therefore selected from the complete VUDS sample galaxies but  keeping only those objects with a CIII] signal to noise  larger than 5, and satisfying a visual inspection such that CIII] is clear of sky subtraction residual or other spectral defect. A total of 43, 31, 30, and 16 spectra are stacked for the weak, medium, strong and very strong emitters, respectively.  Stacked spectra for these different categories are shown in Fig. \ref{stacks_c3}. For the "SFG" sample we do not apply any signal to noise restriction, as the sample is large enough to produce a stable stack.

The spectra show interesting features which change when  EW(CIII]) is increasing: besides the Ly$\alpha$ properties discussed in Sect. \ref{Lya_c3}, we find emission lines of several atomic species which appear and grow stronger.  For the whole SFG population we observe both CIII] and HeII-$\lambda$1640 in emission, as well as a fainter CIV-$\lambda$1549 emission partially suppressed by strong absorption with a P-Cygni profile. Observing both CIII] and HeII in emission is expected as these two lines are ionized under similar conditions as the energy required to  ionize helium is only slightly greater than the energy required to doubly-ionize carbon (24.6 vs. 24.4 eV), and faint HeII emission is frequently observed  in the general population. The HeII emission presents a useful diagnostic for very young as well as more exotic stellar populations \citep{Schaerer2003,Cassata2013} and will be discussed in a forthcoming paper.

\begin{figure*}
   \centering
   \includegraphics[width=18cm,bbllx=1,bblly=230,bburx=591,bbury=630,clip=]{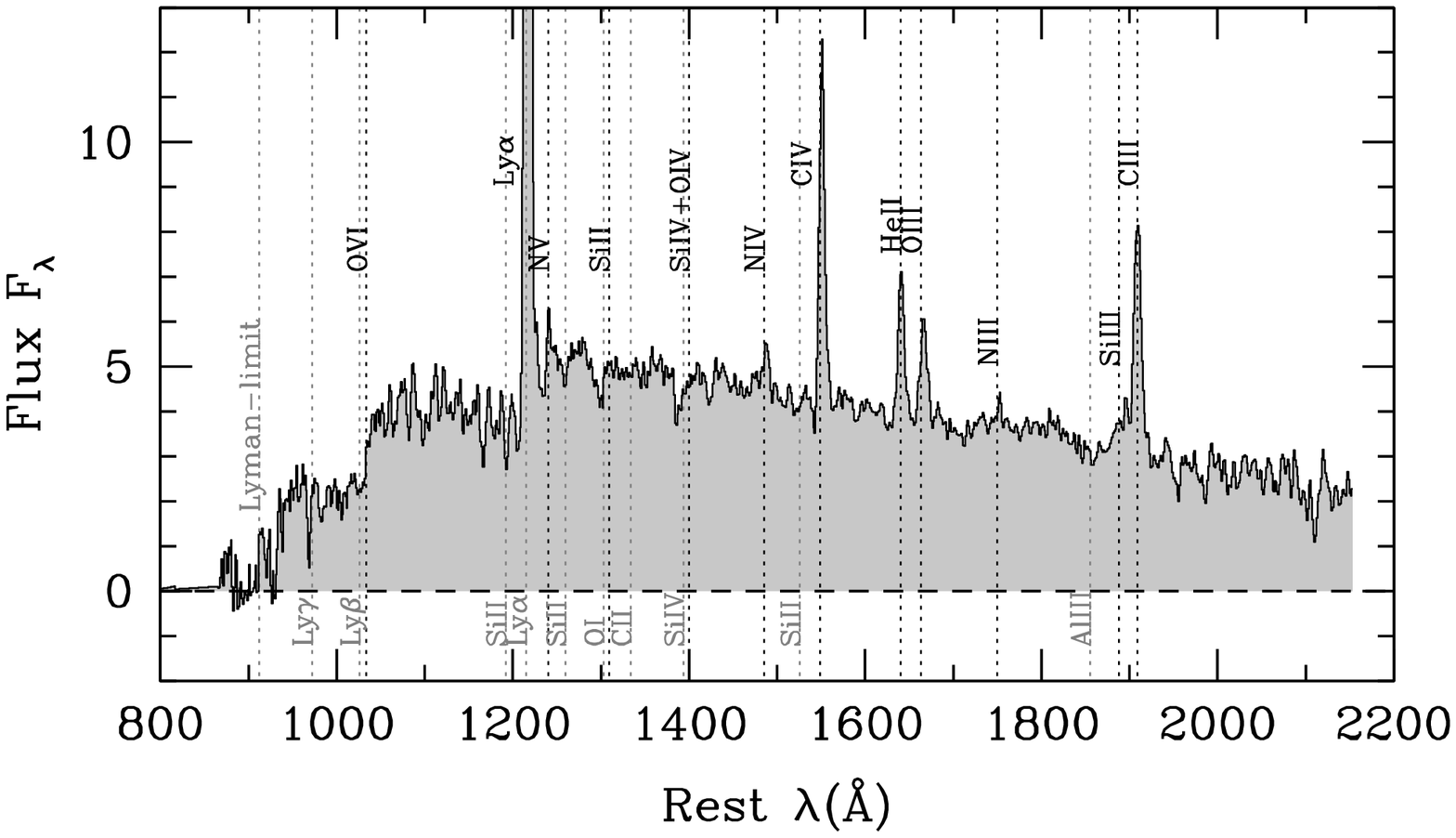}
      \caption{Composite spectrum of 16 CIII]-$\lambda$1908 emitters identified in the VUDS survey at $2<z<3.8$, with CIII] detected in emission with $S/N>5$ and EW(CIII])$>$20\AA. The main emission lines associated to CIII]-$\lambda$1908 ~emission can be identified as: Ly$\alpha$, NV-$\lambda$1240, NIV]-$\lambda$1485, CIV-$\lambda$1549, HeII-$\lambda$1640, OIII]-$\lambda$1663, NIII]-$\lambda$1750, SiIII]-$\lambda$1888. 
}  
         \label{best}
   \end{figure*}

As EW(CIII]) increases we observe in the stacked spectra that HeII and CIV are getting stronger, interstellar medium lines like SiII-$\lambda$1260, OI+SiII-$\lambda$1303, CII-$\lambda$1334, SiV-$\lambda\lambda$1393.76-1402.8, are filling-in, OIII-$\lambda$1663 appears for 5$<$EW(CIII])$<$10, and NIV-$\lambda$1645 and NIII-$\lambda$1750 are both visible for the 10$<$EW(CIII])$<$20 stack. The combined spectrum of the strongest emitters with $EW(CIII])$$\geq$20 strikingly shows a wealth of strong emission lines, as presented in Figure \ref{best}. Besides CIII], we identify the following lines in emission: NV-$\lambda$1240, SiII-$\lambda$1309, SiIV-$\lambda$1403, NIV]-$\lambda$1485, CIV-$\lambda$1549, HeII-$\lambda$1640, [OIII]-$\lambda$1663, NIII]-$\lambda$1750, SiIII]-$\lambda$1888, with a spectrum comparable to that of the UV-selected AGN sample at $z\sim2-3$ from \cite{Hainline2011}. With OIII-$\lambda$1663, SiIII, NIII, NIV and NV requiring energy levels above 35.1, 16.3, 29.6, 47.5, and 77.5 eV, respectively, this is in line with the ionizing field in these galaxies getting stronger. 

We provide measurements of EW and line flux relative to CIII] from the stacked spectra of the different CIII] populations  in Table \ref{stacks_prop}, and in Table \ref{stacks_sfg} for the whole SFG populations at $2<z<3$ and $3<z<4$. We also indicate the slope  $\beta$ of the UV continuum, parametrized as $F_{\lambda} \propto \lambda^{\beta}$, measured from the continuum values between 1400\AA ~and 2100\AA ~rest-frame  \citep{Hathi:16}. To measure these continuum values, we fit the observed continuum with a cubic spline excluding 30\AA ~around known emission and absorption features. We observe that $\beta$ is smaller (continuum steeply declining with wavelength) for the strongest CIII] emitters with $\beta$=$-1.76\pm0.06$, and gradually become flatter for populations with decreasing EW(CIII]), reaching $\beta=-0.92$ for the whole SFG population. This observation corroborates the presence of a stronger UV ionizing field for the strongest CIII] emitters.   


\subsection{Contribution from AGN}
\label{sect_agn}

\begin{figure*}
   \centering
   \includegraphics[width=14cm]{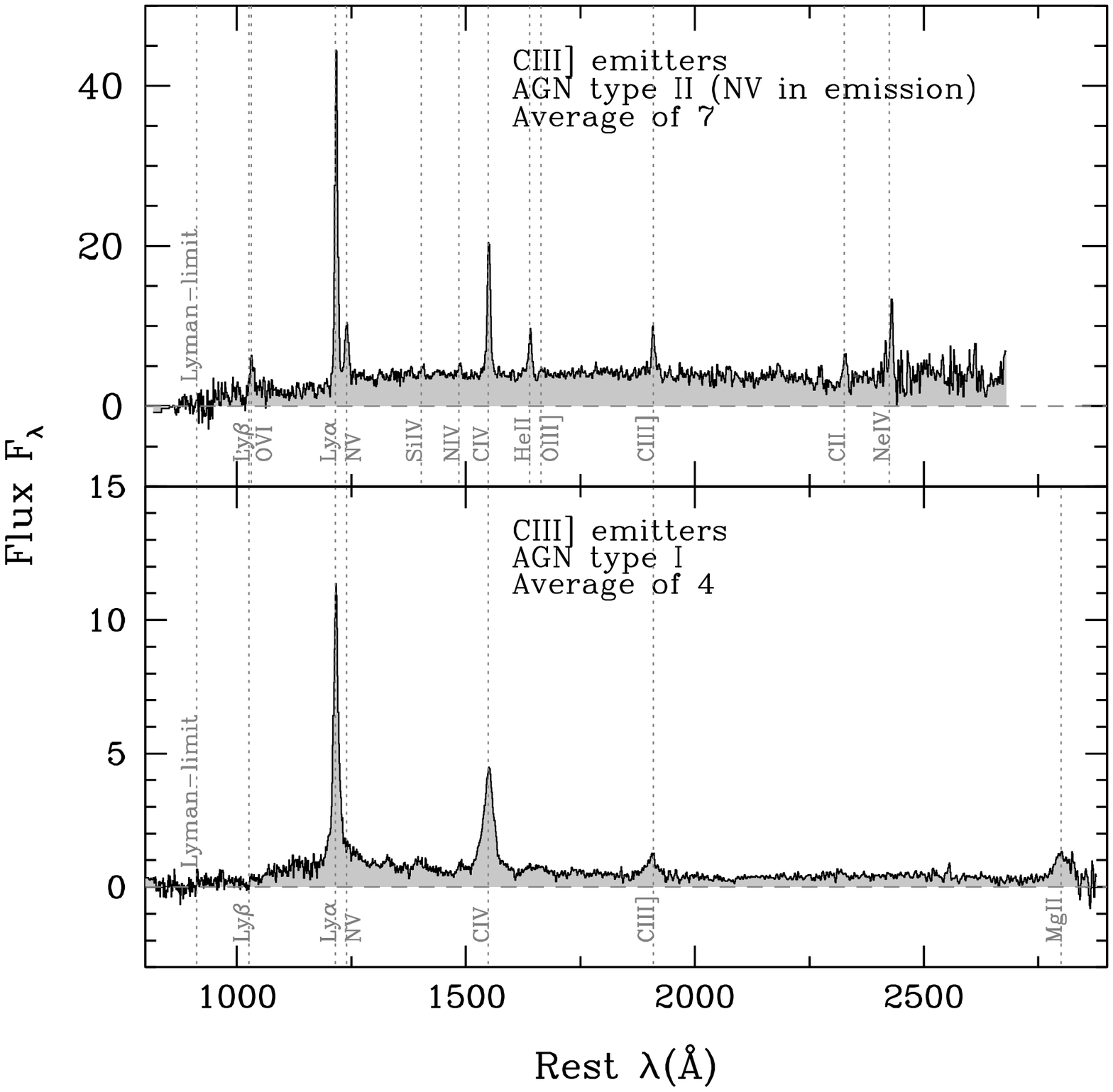}
      \caption{Composite spectrum of 4 type I broad line AGN (bottom panel), and 7 candidate type II narrow line AGN identified from their NV-$\lambda$1240 ~emission (top panel). See text for details on how these AGN are selected. 
}  
         \label{spec_agn}
   \end{figure*}

We make an attempt to identify the contribution of AGN to the SFG population. Four objects among the 3899 sources with $2<z<3.8$ meet the conditions to be classified as broad-line type-I AGN, imposing that one or more of the observed emission lines in the VUDS spectral window have $FWHM>1000$km/s. These objects present spectra similar to broad-line quasar spectra at similar redshifts \citep[e.g.][]{VandenBerk2001}, as can be seen in the stacked spectrum on the bottom panel of Fig.\ref{spec_agn}.  

The NV-$\lambda$1240 ~emission requires a strong ionizing field $\sim$77 eV, and therefore has been proposed as a criterion to identify narrow line type-II AGN \citep[e.g][]{Hainline2011}. As this criterion may not strictly be exclusive of other ionization source for weak NV emission, we search for emitters with EW(NV)$>$10\AA  ~which are most likely type-II AGN, and we identify seven such narrow emission line objects. 

The stacked spectrum of the strong NV emitters is shown in the top panel of Fig.\ref{spec_agn}. We easily identify Ly$\beta$, OVI-$\lambda$1032, Ly$\alpha$, NV-$\lambda$1240, SiIV-$\lambda$1403, NIV-$\lambda$1485, CIV-$\lambda$1549, HeII-$\lambda$1640, OIII]-$\lambda$1664, CIII]-$\lambda$1908, CII]-$\lambda$2326, [NeIV]-$\lambda$2424,  in emission.
The  CIV/CIII] flux ratio for these VUDS narrow-line AGN range from 1.2 to 4.7 with a median CIV/CIII]$=2.7$. This ratio is lower than the ‘class A’ composite spectrum of type II quasars presented in \cite{Alexandroff2013}  with CIV/CIII]$=7.5$. However, it is  higher than the sample of type II AGN in \citet{Hainline2011} for which CIV/CIII]$\sim1.3$,  rather consistent to that of radio-galaxies at similar redshifts \citep{Villar97,Stern1999,Matsuoka2009}, is higher than SFGs which have a typical ratio below unity \citep{Nakajima:2017}. 

Some contribution to the CIV-$\lambda$1549, SiIV-$\lambda$1403 and NV-$\lambda$1240 may be coming from stellar winds  produced by strong star formation. Studying a sample of local analogs to high-z Lyman break galaxies \cite{Heckman2011} indicate that  a broad NV-$\lambda$1240 line, with a P-cygni profile, could arise from stellar winds produced by hot (O) stars. Still only a few galaxies have CIV/CIII values comparable to even the lowest values among the NV-detected galaxies, making a pure star formation interpretation unlikely to be correct at these low redshifts.

Interestingly two out of seven strong NV emitters show weak or absent Ly$\alpha$ emission (see object 520391642 in Fig.\ref{strong_c3}). These two objects present a red UV continuum ($\beta$ slope $+2.3$ and $+0.8$) and may therefore have dusty ISM conditions which are preventing Ly$\alpha$ to escape. The AGN type II stack has a much redder continuum slope $\beta=+0.2\pm0.05$ compared to the CIII]-selected star forming populations discussed in Sect.\ref{c3_pop} with $\beta>-1.6$. 
These results are in line with those presented by \cite{Hainline2011}. This indicates that the internal extinction is stronger in the AGN population, and that a combination of a dust-obscured AGN and star formation component may combine to produce the observed spectra. 

We note that the AGN sample we have identified here based on NV are almost certainly, as we shall see later, a small subset of the total AGN population. The criteria to identify type II AGN from UV-rest spectra vary from sample to sample in the literature, and we point out the necessity to perform a quantitative spectral line analysis rather than an empirical classification.  This is developed in \citet{Nakajima:2017} where various diagnostic diagrams using UV emission lines and based on photoionization models are discussed, and in particular pointing to the need of using  different line ratios to separate star formation from AGN. \citet{Nakajima:2017} claim that the population of strong and very strong CIII] emitters presented here must include AGN to explain the high EW(CIII]) and the location of these galaxies in diagnostic diagrams using observed UV line ratios. This is further discussed in Sect.\ref{properties} and \ref{agn_quench}.


We searched for individual X-ray counterparts of CIII] emitters using Chandra observations in the CDFS \citep[the 7 Ms catalog of][]{Luo:2017} , in COSMOS \citep[the Legacy catalog of][]{Civano:2016,Marchesi:2016} , and the XMM data in the VVDS-02h field \citep[the XMM-LSS catalog of][]{Chiappetti:2013}, with flux limits L$_X$(2-10keV) ranging from $3\times10^{42}$, to $3\times10^{44}$ erg/s at z$\sim$3. We found a total of 8 X-ray counterparts, one in CDFS (id \#530029221), 6 in COSMOS (ids \#510080080, 510788005, 511707398, 510793659, 510164473, and 5121237451) and one in the VVDS-02h (id \#520426550) \citep[see also ][]{Talia2016}. These sources have X-ray luminosity in the range 43.8$<$log(L$_x$(2-10keV)$<$44.5. All of these objects are identified as AGN following the NV analysis described above, except one (id \#510788005) which is identified in the strong CIII] emitters list. We then performed X-ray stacking analysis on the list of strong and very strong emitters in the COSMOS field using the maps and source catalog of \citet{Civano:2012}. We used the the CSTACK tool to separately stack the sources in the very strong and strong CIII] emitters lists after excluding the two objects with an individual detection. 
In the stack of galaxies in the very strong sample the corresponding mean count rate per galaxy implies logL$_x$(2-10keV)$\sim $42.6 erg/s at z$\sim$3, with a noise level in the map of logL$_x$(2-10keV)$\sim$42.7 erg/s, meaning no significant detection. Conversely, in the strong sample there is a $2\sigma$ detection with a count rate implying logL$_x$(2-10keV)$\sim$42.9 erg/s, with a noise level in the map of logL$_x$(2-10keV)$\sim$42.6 erg/s. The current X-ray imaging is then not deep enough to firmly confirm the presence of an X-ray source. However, the tentative 2$\sigma$ detection indicates that  X-ray imaging is consistent with the very strong and strong CIII] emitting galaxies hosting low-luminosity AGN with X-ray luminosity as observed on the faint end of the AGN luminosity function at these redshifts \citep{Georgakakis:2015}. 



\section{Lyman$-\alpha$ properties of CIII] emitters}
\label{Lya_c3}

\begin{figure*}
   \centering
   \includegraphics[width=14cm]{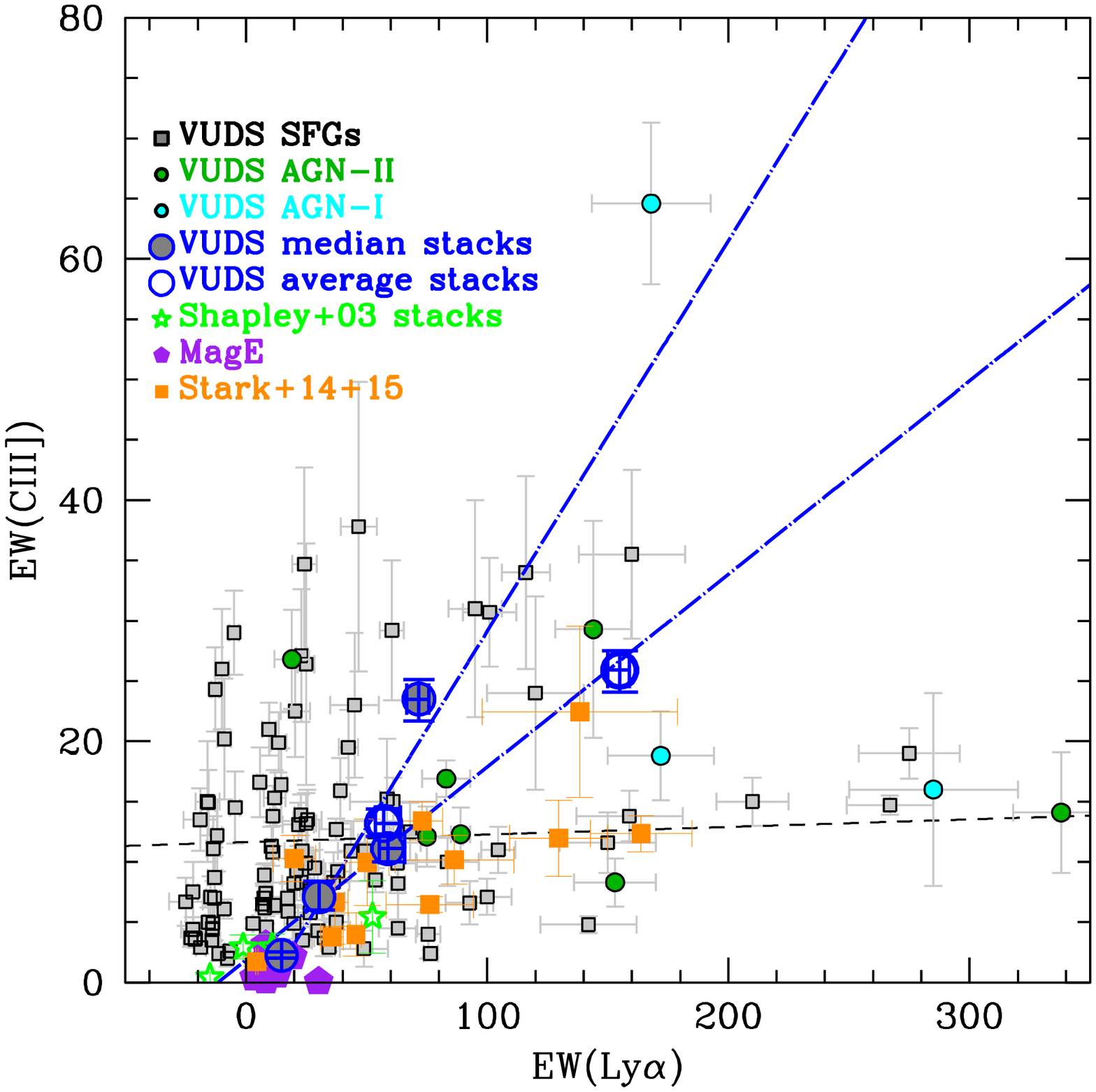}
      \caption{Rest-frame equivalent width of CIII]-$\lambda$1908 EW(CIII]) as a function of EW(Ly$\alpha$). The least square fit to the individual data points (black squares) is represented by the black dashed line. The individual galaxies show only a weak correlation, significant at about the 3$\sigma$ level, with a large dispersion , with a large dispersion. Galaxies with strong Ly$\alpha$ and weak CIII] are observed, as well as galaxies with strong CIII] and Ly$\alpha$ in absorption. The EW(CIII]) versus EW(Ly$\alpha$) points for the median-stacked spectra of the different categories defined in Sect.\ref{c3_pop} are indicated as blue circles filled in gray, while values for the average stacks in Tab.\ref{stacks_prop} are indicated as the open blue circles. The correlation between  EW(CIII]) and EW(Ly$\alpha$) appears as very strong ($r_P=0.99$, $r_S=1$), hiding the wide range of properties of CIII] vs. Ly$\alpha$. As a comparison with the literature, we plot the observed values from the different Ly$\alpha$ samples of \citet{Shapley2003} (green stars, average for Ly$\alpha$ samples), \citet{Rigby2015} (purple diamonds), and \citet{Stark2014,Stark2015} (orange squares).
}  
         \label{lya_c3}
   \end{figure*}

We investigate the relationship between CIII] and Ly$\alpha$ emission in Figure \ref{lya_c3} using the sample of CIII] emitters with S/N$>$5. Looking at the relation between CIII] and Ly$\alpha$ from the spectral stacks (Tab.\ref{stacks_prop}), we find a strong correlation with a Pearson correlation coefficient $r_P=0.99$ and Spearman rank correlation coefficient $r_S=1$ ($r_P$ or $r_S$ equal to one indicate a perfect one to one correlation).  However, from the individual emitters, while we observe  a trend for the strongest CIII] emitters  to also tend to be strong Ly$\alpha$ emitters with a correlation significant at about the 3$\sigma$ level (Student t-test), this correlation has a large dispersion. 
This correlation further weakens when restricting the sample of CIII] emitters to EW(Ly$\alpha$)$<$50\AA ~with $r_P=0.2$, similarly to that noted by \citet{Rigby2015}. 

The analysis of individual emitters demonstrates that the presence of Ly$\alpha$ in emission  increases the probability of detecting CIII], but that strong Ly$\alpha$ emission  does not necessarily mean strong CIII] emission, or vice-versa. There are indeed numerous examples of galaxies with CIII] in emission and no or weak Ly$\alpha$ emission (some are presented in Fig.\ref{strong_c3} and Fig.\ref{weak_c3}). We reach similar conclusions when looking at the CIV-$\lambda$1549 ~line versus Ly$\alpha$ emission. 

This behavior is somewhat expected as the radiative transfer of Ly$\alpha$ or CIII] photons is  markedly different.  As a striking example we point to object 520391642 (second from top in Figure \ref{strong_c3}) which shows strong CIII] (EW$>$20), as well as CIV, HeII or NV, but only weak Ly$\alpha$ emission. 
Our results compare well with the literature once the large spread of the relation between EW(CIII]) and EW(Ly$\alpha$), the selection, and the size of different samples, is taken into account. Looking at the average properties from stacks, our results agree well with \citet{Shapley2003}. VUDS galaxies span a broad range of properties, including galaxies with and without Ly$\alpha$ in emission, expanding from smaller and possibly biased samples for which a stronger correlation between Ly$\alpha$ and CIII] is reported \citep{Stark2014,Stark2015,Rigby2015,Guaita2017}. 

We find that galaxies with EW(Ly$\alpha$)$>$25 (the definition of a Ly$\alpha$ emitter, LAE), have a 42\% and 20\% chance to be emitting CIII] and CIV at an appreciable level (EW$>$3\AA), respectively. For those galaxies emitting both Ly$\alpha$ and CIII] at these levels, the median EW(CIII]) is $7.7$\AA ~as compared to a median EW(CIV) of $3.4$\AA ~for the average LAE emitting CIV.


\section{Distribution of EW(CIII])}
\label{distrib}

To answer the question of how frequently CIII] emission is observed in  UV-selected galaxies, we seek to establish the fraction of CIII] emitters in the VUDS sample as a function of EW(CIII]). While it is  straightforward to identify strong CIII] emitters it is more difficult to separate weak emitters from noise. We therefore proceed with a statistical analysis of the distribution of CIII] emission for the 3899 VUDS galaxies with $2<z<3.8$. 

If CIII] emission is absent and in the presence of noise, the line measurement resulting from the process presented in Sect. \ref{c3_measurement} could result on either a false positive or  negative EW.  As CIII] is only expected in emission, we use the distribution of EW measurements indicating absorption as an estimate of how many  CIII] emission lines could be produced from noise spikes, under the assumption of a Gaussian noise distribution. We compute the distribution for EW(CIII])$<$0, make the EW of this distribution positive (emission), and subtract this distribution from the distribution of EW(CIII])$>$0 to statistically correct the distribution of EW(CIII]) for false CIII] emission measurements and obtain the EW distribution for CIII] emitters. For each EW bin we compute the associated error from the quadratic sum of the observed numbers of emitters and non-emitters.  

\begin{figure*}
   \centering
   \includegraphics[width=14cm,bbllx=1,bblly=270,bburx=591,bbury=630,clip=]{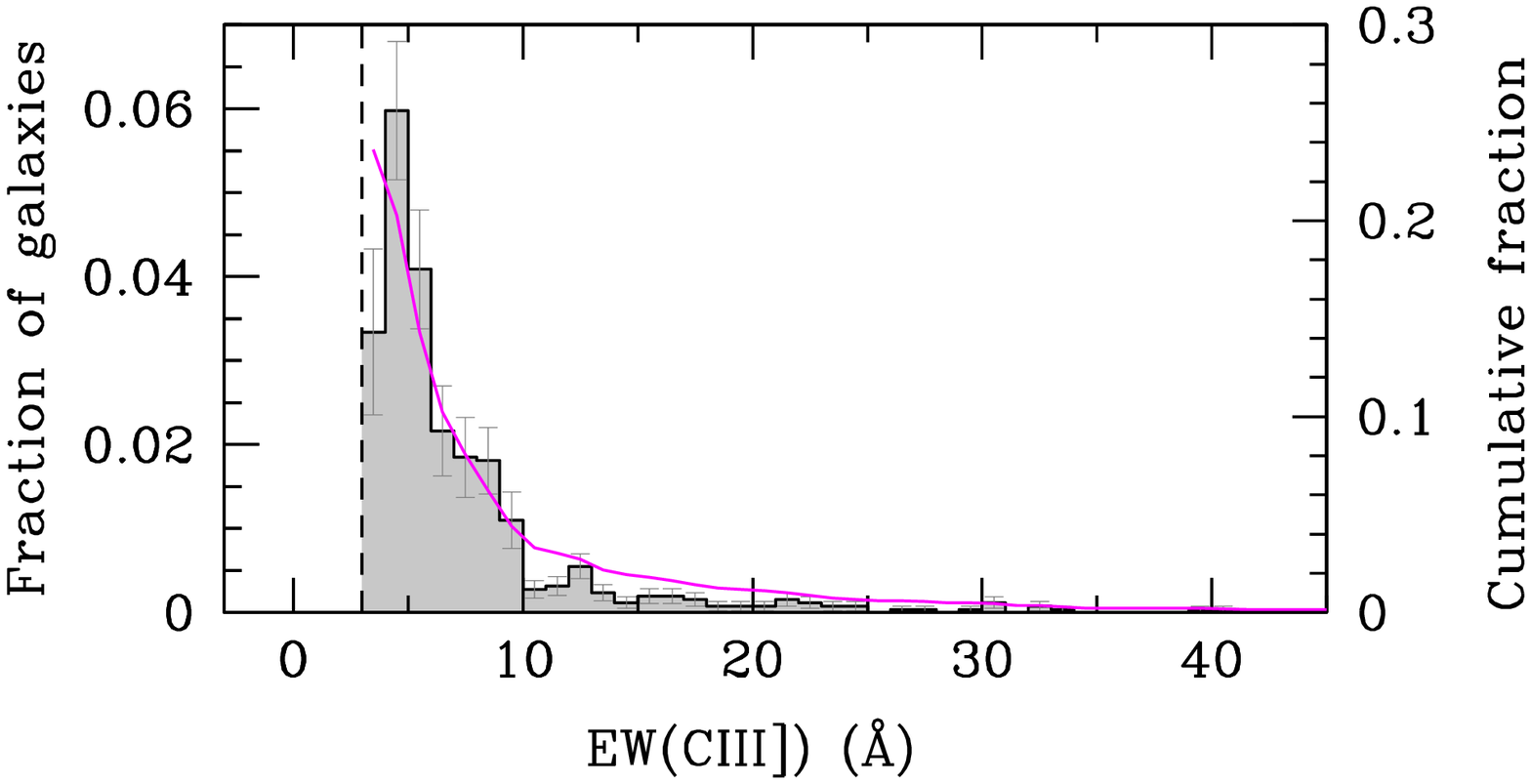}
      \caption{Distribution of CIII]-$\lambda$1908 equivalent width EW(CIII]) in the UV-selected star-forming population at $2<z<3.8$. The shaded histogram shows the fraction of galaxies with a given EW(CIII]) in bins of $\delta$EW=1\AA ~(left Y-axis), with Poisson uncertainties from the population statistics indicated as error bars. The cumulative fraction of galaxies with EW(CIII]) above a given EW(CIII]) limit is drawn as the continuous magenta line (values on the right Y-axis), relative to  the total population of SFGs at that redshift. About 24\% of SFGs with $2<z<3.8$ present CIII]-$\lambda$1908\AA ~emission with EW(CIII])$\geq$3\AA.  }  
         \label{dist_c3}
   \end{figure*}

The resulting rest-frame EW(CIII]) distribution is shown in Figure \ref{dist_c3}, presented as the fraction of emitters with a given EW(CIII]) over the full population of UV-selected star-forming galaxies, both differential and cumulative. The fraction of emitters decreases from a peak at low EW down to the largest narrow line emission galaxies with $EW(CIII])\sim40$\AA, as expected. We find that the population of star-forming galaxies contains $4.0\pm1.2$\% of strong emitters with $EW(CIII])>10$\AA, including $1.2\pm0.4$\% of very strong CIII] emitters with $EW(CIII])>20$\AA, and $20\pm3.1$\% of weak emitters $3<EW(CIII])<10$\AA.   The line emission distribution shows that CIII] emission is present above EW(CIII])=3\AA ~for less than one fourth of UV star-forming galaxies at $z\sim3$.

Besides Ly$\alpha$, the CIII] line remains the most frequent emission line observed at EW$>$3\AA. By comparison a similar analysis shows that about 10\% of the galaxies emit CIV at a level EW(CIV)$>$3\AA ~sufficient to claim detection from VUDS spectra. 


\section{Properties of CIII] emitters}
\label{properties}

\subsection{L$_{NUV}$, M$_{star}$, SFR, morphology, of CIII] emitters}
\label{all_prop}

\begin{figure*}
   \centering
   \includegraphics[width=14cm]{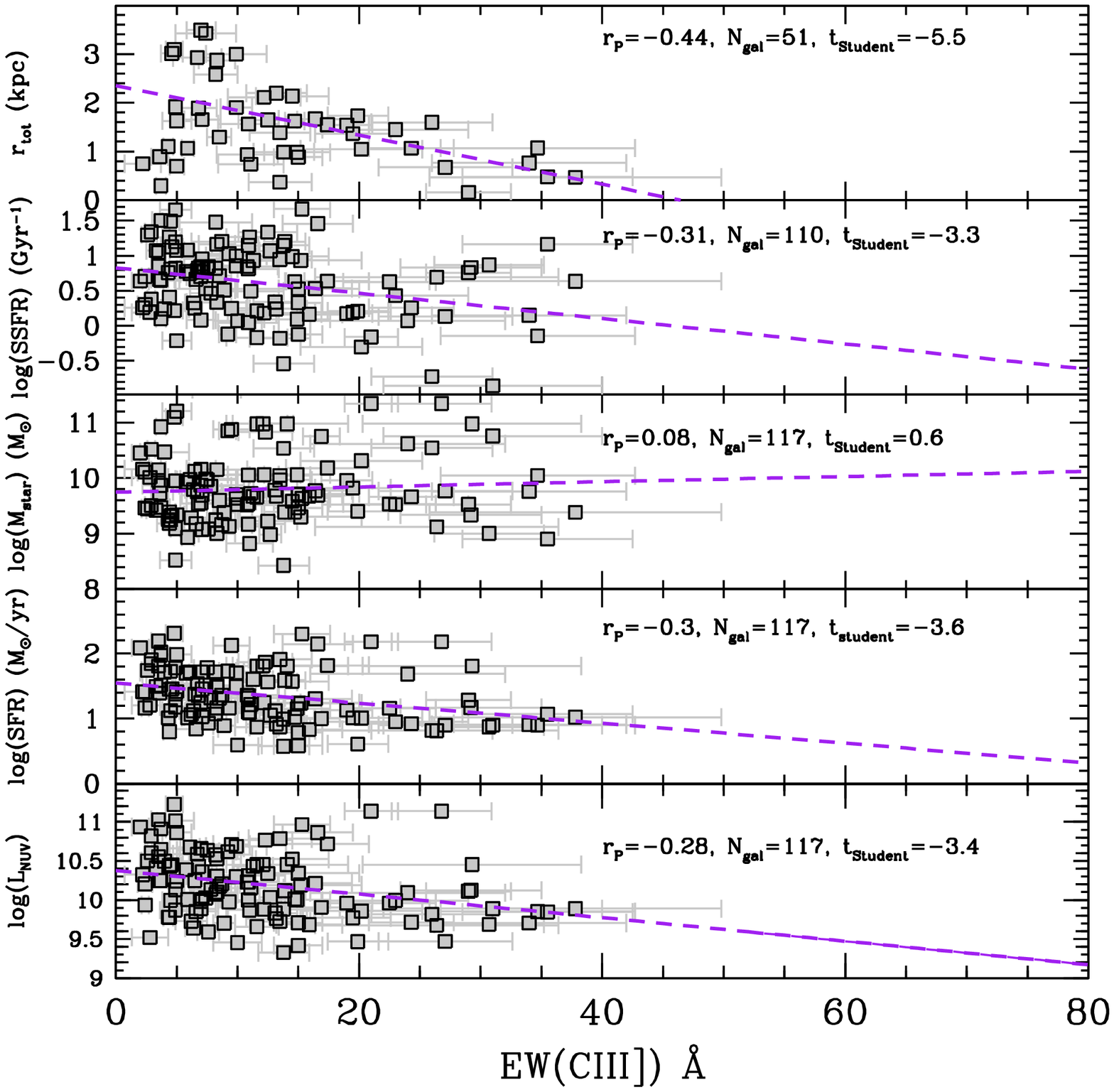}
      \caption{Properties of CIII] emitters, from bottom to top: L$_{NUV}$, SFR, M$_{star}$, SSFR, and total size  properties of CIII] emitters as a function of EW(CIII]). A least square fit to the data is shown as the purple dashed lines, and the Pearson correlation coefficient $r_P$, number of objects, and the significance of the correlation (Student t-test), are indicated on the upper right corner of each panel. Note that the r$_{tot}$ measurements are limited to the COSMOS and ECDFS fields where HST imaging is available \citep{Ribeiro2016}. 
}  
         \label{c3_prop}
   \end{figure*}

In this section we compare the population of CIII] emitters to the general properties of the star-forming galaxy population. We present the relation between the rest-frame near UV luminosity L$_{NUV}$, the SFR, M$_{star}$, SSFR, and total size r$_{tot}$, as a function of EW(CIII]) in Fig.\ref{c3_prop}, including the significance of the correlation using the Pearson correlation coefficient r$_P$ and  Student t-test significance.

We find that the L$_{NUV}$ luminosity is decreasing with increasing EW(CIII]), from a median \~L$_{NUV}$=10.2 in 5$<$EW(CIII])$<$20 to \~L$_{NUV}$=9.9 for EW(CIII])$\geq$20, with r$_P=-0.28$, a correlation significant at the 3.4$\sigma$ level. If this trend is maintained to higher redshifts, one may expect to observe a higher fraction of CIII] emitters in lower luminosity galaxies into the reionisation epoch than reported in Sect.\ref{distrib}. 

The SFR is also  decreasing with increasing EW(CIII]) (r$_P$=0.3, significant at 3.6$\sigma$), 
with a decrease of $\sim$20\% for emitters with EW(CIII])$\geq$20 compared to 5$<$EW(CIII])$<$20. With EW(CIII])  dependent on the ionizing field, and the strongest CIII] emitters partially powered by AGN \citep{Nakajima:2017}, this relation may be due to the quenching of star-formation. This is discussed in Sect.\ref{main_seq}. 

There is a weak trend for the strongest CIII] emitters to have  higher stellar masses than the weaker emitters, but this is not significant (r$_P=0.08$, 0.6$\sigma$). The trend in SFR and M$_{star}$ combine for a weak dependency of SSFR with EW(CIII]) (r$_P=-0.31$, 3.3$\sigma$). There is no apparent trend with EW(CIII]) of the dust extinction from the E(B-V) computed from SED fitting.


A visual morphological classification is performed on each of the CIII] samples (Tab.\ref{morph}), using the HST F814W images available in the COSMOS and ECDFS fields \citep[for details on the imaging data on the VUDS sample, see][]{Ribeiro2016}. The very strong CIII] emitters have a morphology dominated by compact  galaxies \citep[r$_{100}$$<$1 kpc, see][]{Ribeiro2016} and galaxy pairs with separation less than 20 kpc \citep[see][]{Ribeiro2017}. The population of strong CIII] emitters shows a high fraction of compact galaxies, with an  equal fraction of pairs. For the medium and weak emitters, the fraction of compact galaxies is low, the pair fraction remains similar to the other CIII] samples, but clumpy galaxies and those extended and symmetric dominate. It therefore appears that CIII]  is preferentially emitted in compact and interacting systems, as already pointed out in the smaller VUDS sub-sample of \citet{Amorin2017}. This is further supported by the distribution of sizes as a function of EW(CIII]) presented in Fig.\ref{c3_prop}, with sizes clearly decreasing with increasing EW(CIII]) (with r$_P$=$-0.44$, a correlation significant at the 5.5$\sigma$ level).


\begin{table}
\caption{Visual classification of CIII] emitters morphology}
\label{morph}    
\begin{tabular}{|c|c|c|c|}       
\hline             
 Sample                          &  Compact   & Pairs & Clumpy \& Extended \\ 
                                      &   r$_T$$<$1kpc$^1$          &  r$_p$$<$20kpc$^2$ &         \\  \hline  
0$<$EW(CIII])$<$10\AA     &       6\%                          &  40\%                      &    54\%  \\ \hline
10$<$EW(CIII])$<$20\AA   &     39\%                          & 39\%                      &     22\%   \\ \hline
EW(CIII])$>$20\AA              &    25\%                          & 44\%                       &    31\%   \\ \hline                           
\end{tabular}
\begin{center}
 \begin{list}{}{}
 \item[$^{\mathrm{1}}$ Total radius \citep{Ribeiro2016}]
 \item[$^{\mathrm{2}}$ Transverse pair separation]
 \end{list}
\end{center}
\end{table}

\subsection{Outflow velocities}

\begin{figure*}
   \centering
   \includegraphics[width=12cm]{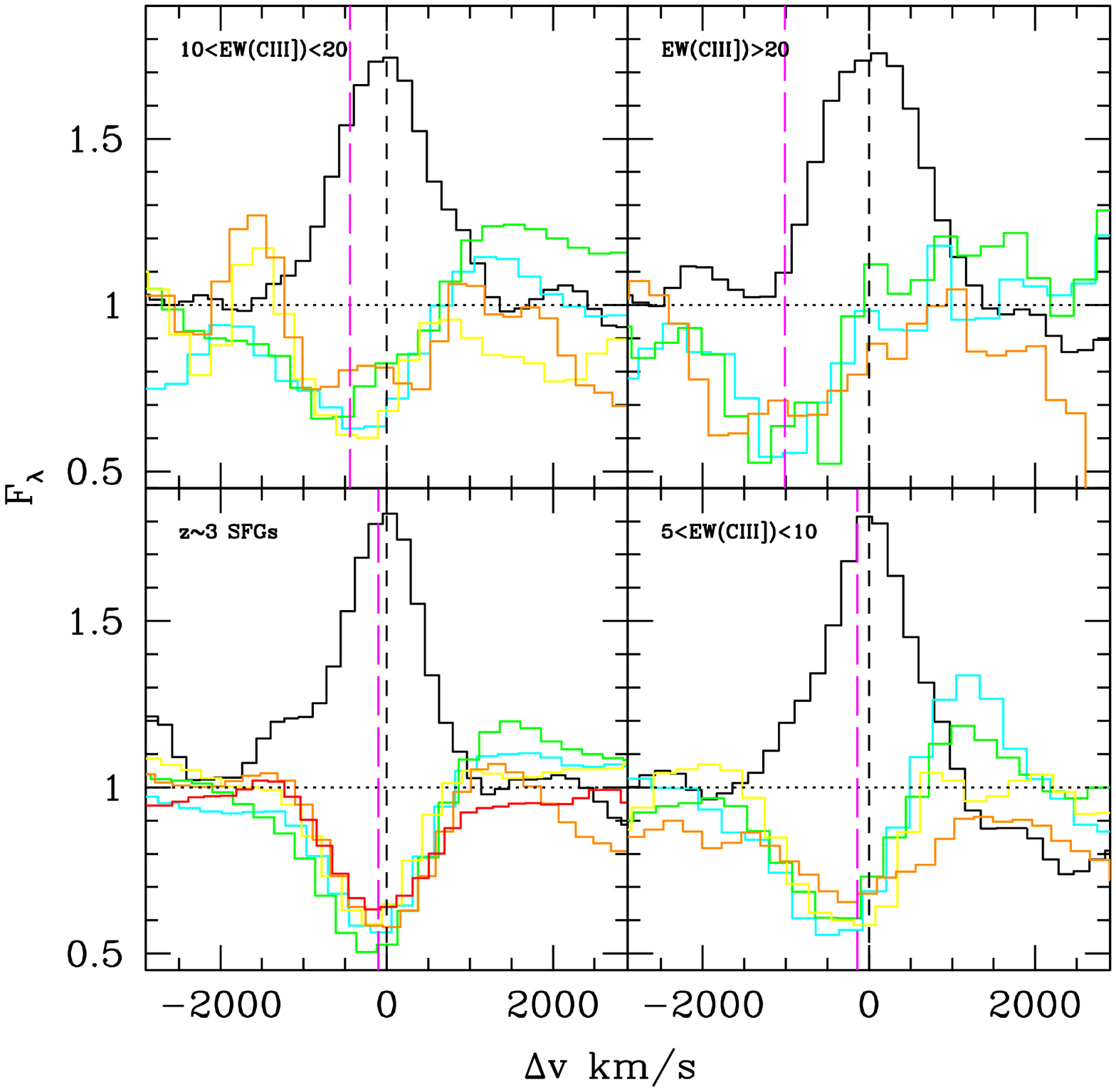}
      \caption{Relative velocity of the main ISM absorption lines with respect to CIII]-$\lambda1908$\AA ~taken at systemic velocity, for the SFG population and samples with increasing EW(CIII]).  CIII] is in black, and depending on the sample SiII-$\lambda$1260 is drawn in  cyan, SiII-$\lambda$1303 in green, CII-$\lambda$1334 in yellow, 1526 in orange, and 1670 in red. 
}  
         \label{velocity}
   \end{figure*}

\begin{figure*}
   \centering
   \includegraphics[width=10cm]{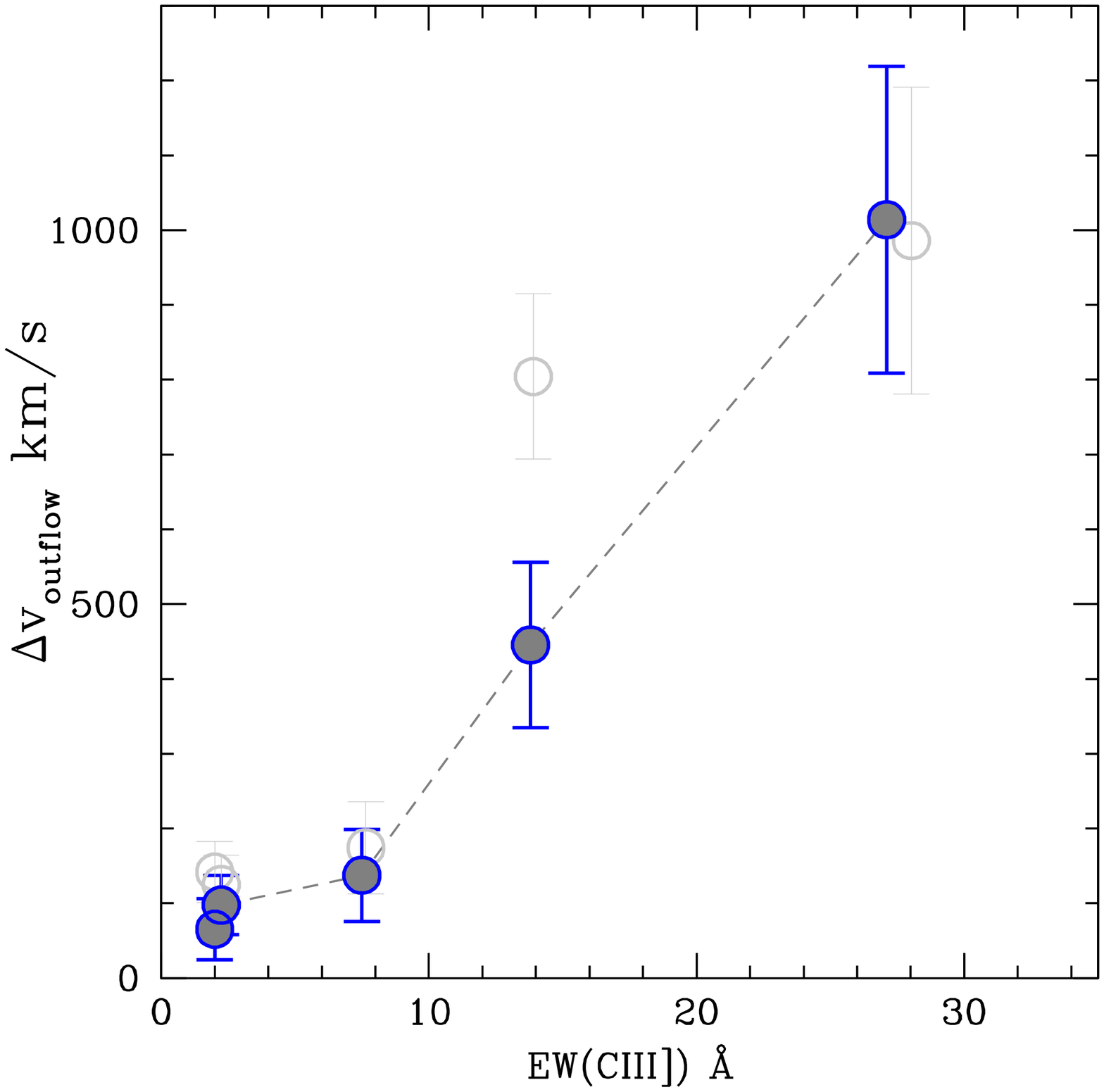}
      \caption{Outflow velocity of ISM lines as a function of EW(CIII]) (filled symbols: velocity from the median values of the ISM lines, open symbols: mean values) 
}  
         \label{velocity_c3}
   \end{figure*}

The velocity of ISM lines compared to the systemic velocity is presented in Fig.\ref{velocity} as measured from the stacked CIII] spectra, and the median velocity derived from the individual line measurements is presented in Fig.\ref{velocity_c3}. We set the systemic velocity on the  CIII] line, and verified that this agrees with photospheric absorption lines CIII]-$\lambda$1176 and SV-$\lambda$1501, as well as with HeII-$\lambda$1640 emission \citep[see ][]{Talia2016}. To measure the velocity of the ISM we use the ISM absorption lines in Table 2 of \citet{Talia2016}, except the FeII-$\lambda$1608 line which is measured to be compatible for being at rest and may be contaminated by a blend of absorption features (FeIII, AlIII, NII) as pointed out by \citet{Talia2016}. We also exclude the CIV-$\lambda$1549 line, which is a combination of emission and absorption. For the whole SFG population at $2<z<3$ and $3<z<4$ and the CIII] emitters with 5$<$EW(CIII])$<$10\AA ~we detect and use the above lines to compute the median velocity difference $\Delta V=\delta \lambda / \lambda \times c$.  For  5$<$EW(CIII])$<$10\AA ~emitters we do not use SiV-$\lambda$1402.77 and AlIII -$\lambda$1854.72,1862.79 only weakly detected. For EW(CIII])$>$20 we use  SiII-$\lambda$1260, SiIV-$\lambda$1393.76, 1402,77, SiII-$\lambda$1526.71, and AlIII-$\lambda$1854.72,1862.79, with the other lines not detected in the stack of this smaller sample. At the spectral resolution used for the VUDS sample (Sect. \ref{survey}), the instrumental resolution is $\sim$1000km/s, allowing for an accuracy in relative velocity measurements of $\sim$250km/s  \citep{LeFevre:15}, and when all ISM lines are well detected as in the whole SFG samples at $2<z<3$ and $3<z<4$, the velocity dispersion of line velocity measurements is $\sigma_{V}(ISM)\sim40-50$km/s.

The resulting relative velocity measurements are quite striking (Fig.\ref{velocity_c3}): we find  low velocity outflows with 50 to 150 km/s for the whole SFG populations at $2<z<3$ and $3<z<4$ and medium CIII] emitters, while the outflow velocity increases to $445\pm110$ km/s for the strong 10$<$EW(CIII])$<$20 emitters, and reaches $1014\pm205$ km/s for the very strong emitters with EW(CIII])$>$20, well beyond the escape velocity of such galaxies \citep{Cimatti:2013}, and comparable to what is observed in SFGs with AGN at $z\sim1-2$ \citep[e.g.][]{genzel2014}.  As star formation does not seem to be capable to produce outflow velocities beyond $\sim$600 km/s even for strong starbursts with maximally efficient winds \citep{Thacker:2006}, we conclude that the outflows reported here for the stronger CIII] emitters are produced by an AGN. This is further discussed in the following sections.

\subsection{CIII] emitters in the SFR vs. M$_{star}$ plane}
\label{main_seq}

\begin{figure*}
   \centering
   \includegraphics[width=14cm,bbllx=1,bblly=220,bburx=591,bbury=630,clip=]{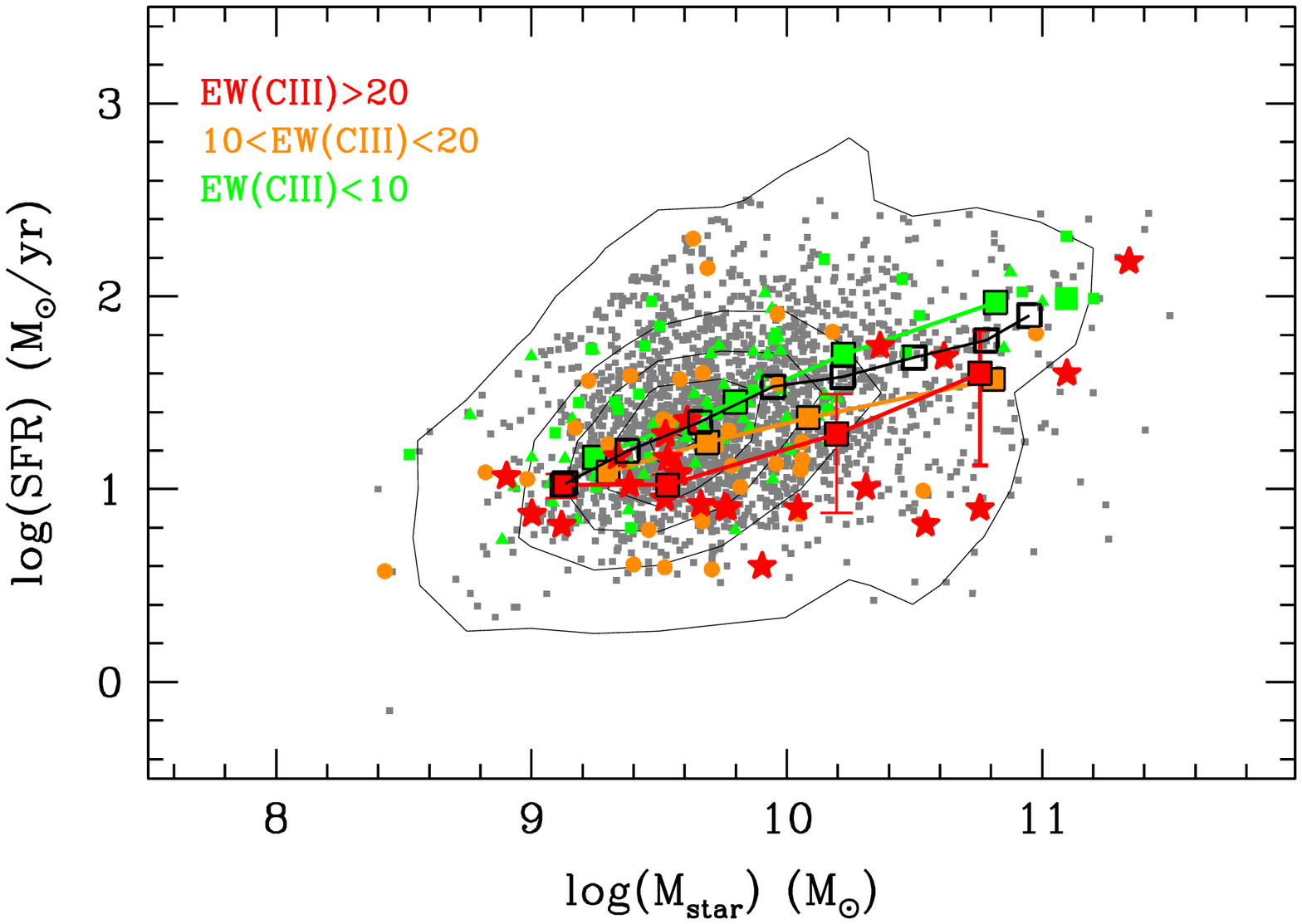}
      \caption{Distribution of CIII] emitters with different EW(CIII]) in the M$_{star}$-SFR plane. The CIII] emitters with different EW are identified by the different colored symbols. The parent sample of star-forming galaxies at redshifts $2<z<3.8$ in VUDS is identified as grey symbols with their distribution identified by the contours, and the main sequence of VUDS galaxies at $z\sim3$ is indicated by black open squares connected by a thick line.  We note that both the strong and very strong CIII] emitters are on average below the MS, with median values identified by the connected orange and red squares, respectively. The lower SFR of these strong CIII] emitters compared to the MS is possibly resulting from a strong star formation quenching produced by AGN  feedback in the host star-forming galaxies.
}  
         \label{ms_c3}
   \end{figure*}

\begin{figure*}
   \centering
   \includegraphics[width=14cm]{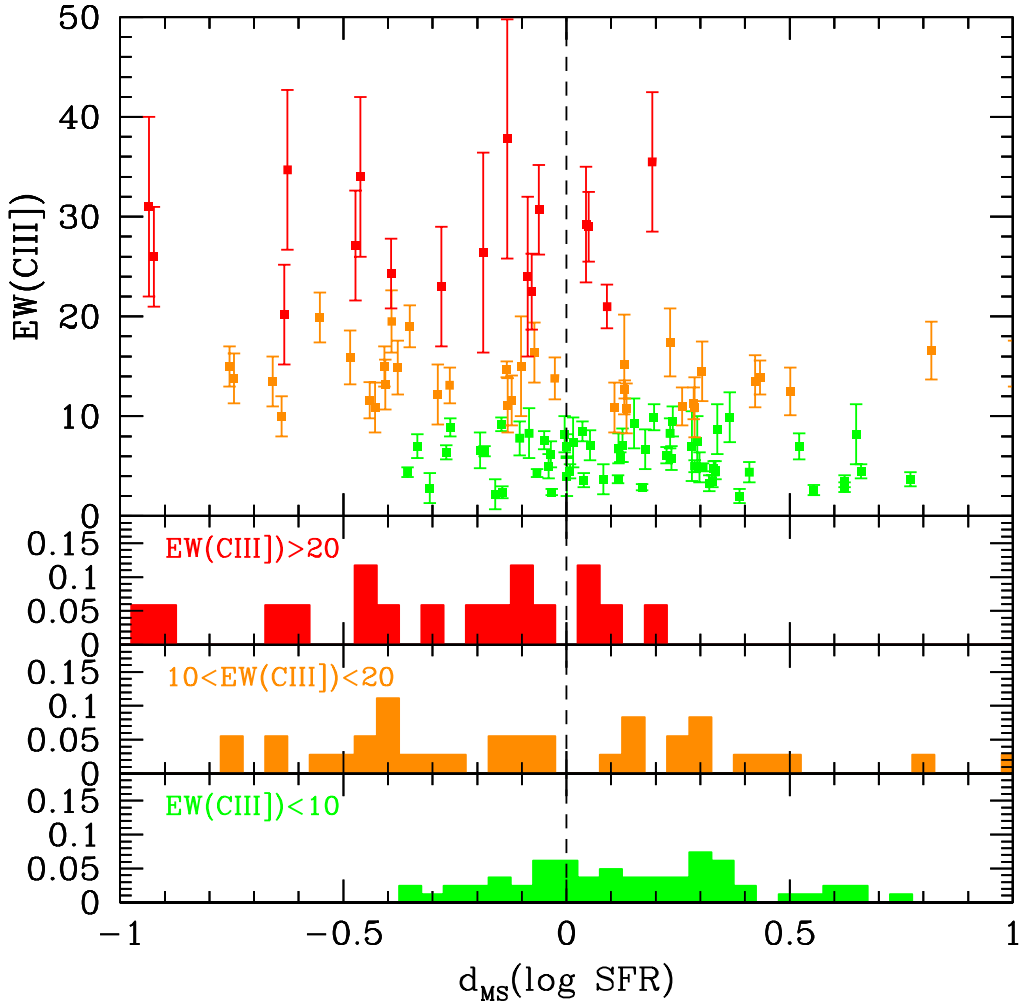}
      \caption{{\it (bottom three panels)} Distribution of distances in SFR from the main sequence for CIII] emitters with increasing EW(CIII]); {\it (top panel)} Distance to the main sequence for individual CIII] emitters. As EW(CIII]) increases the distance to the main sequence increases.
}  
         \label{distance_c3}
   \end{figure*}

The position of CIII] emitters with respect to the main sequence (MS) of star-forming galaxies in the stellar mass vs. SFR plane is presented in Fig.\ref{ms_c3}. The strong and very strong emitters show a trend of being located below the main sequence (MS) observed in the same redshift range \citep{Tasca2015}. This is particularly evident for the very strong emitters with EW(CIII])$>$20, which are mostly below the MS.  
For this population it is quite striking to observe galaxies with low or even very low SFR, but with very strong CIII] emission. 
In Fig.\ref{distance_c3} we present the distribution of the difference between the SFR of galaxies and the main sequence identified in Fig.\ref{ms_c3}. The mean SFR difference increases from d$_{MS}(SFR)=+0.14\pm0.03$ (dex, above the MS) for emitters with 0$<$EW(CIII])$<$10\AA, to d$_{MS}(SFR)=-0.10\pm0.08$ for 10$<$EW(CIII])$<$20\AA, to d$_{MS}(SFR)=-0.27\pm0.07$ (below the MS)  for EW(CIII])$>$20\AA. The total shift in SFR of d$_{MS}(SFR)=-0.41$ dex between the weak CIII] emitters and the very strong emitters is significant at more than a 5$\sigma$ level. 

The star formation of the weaker CIII] emitters appears to be higher than for the bulk of the SFGs. While the whole SFG population has a median EW(CIII])$\sim$2\AA ~(Tab.\ref{stacks_prop}), the population of weak  and medium emitters have EW(CIII])=3.4\AA ~and 8.0\AA, respectively. Producing 3$<$EW(CIII])$<$10\AA ~in the absence of an AGN requires enhanced star formation consistent with this population being above the MS \citep{Nakajima:2017}. 

The distance to the MS of the strong and very strong CIII] emitters is more puzzling, as the distance to the MS of these populations is increasingly below the MS, with a star formation rate about half that on the MS. A possible interpretation is that the star formation for a fraction of the strong emitters and a majority of the very strong emitters is subject to quenching of star formation. Some of the very strong emitters are more than a factor of 10 below the MS, which means that the quenching must be very effective, while, at the same time, maintaining a very strong ionizing field needed to produce EW(CIII]) in excess of 20\AA.

\begin{figure*}
   \centering
   \includegraphics[width=14cm]{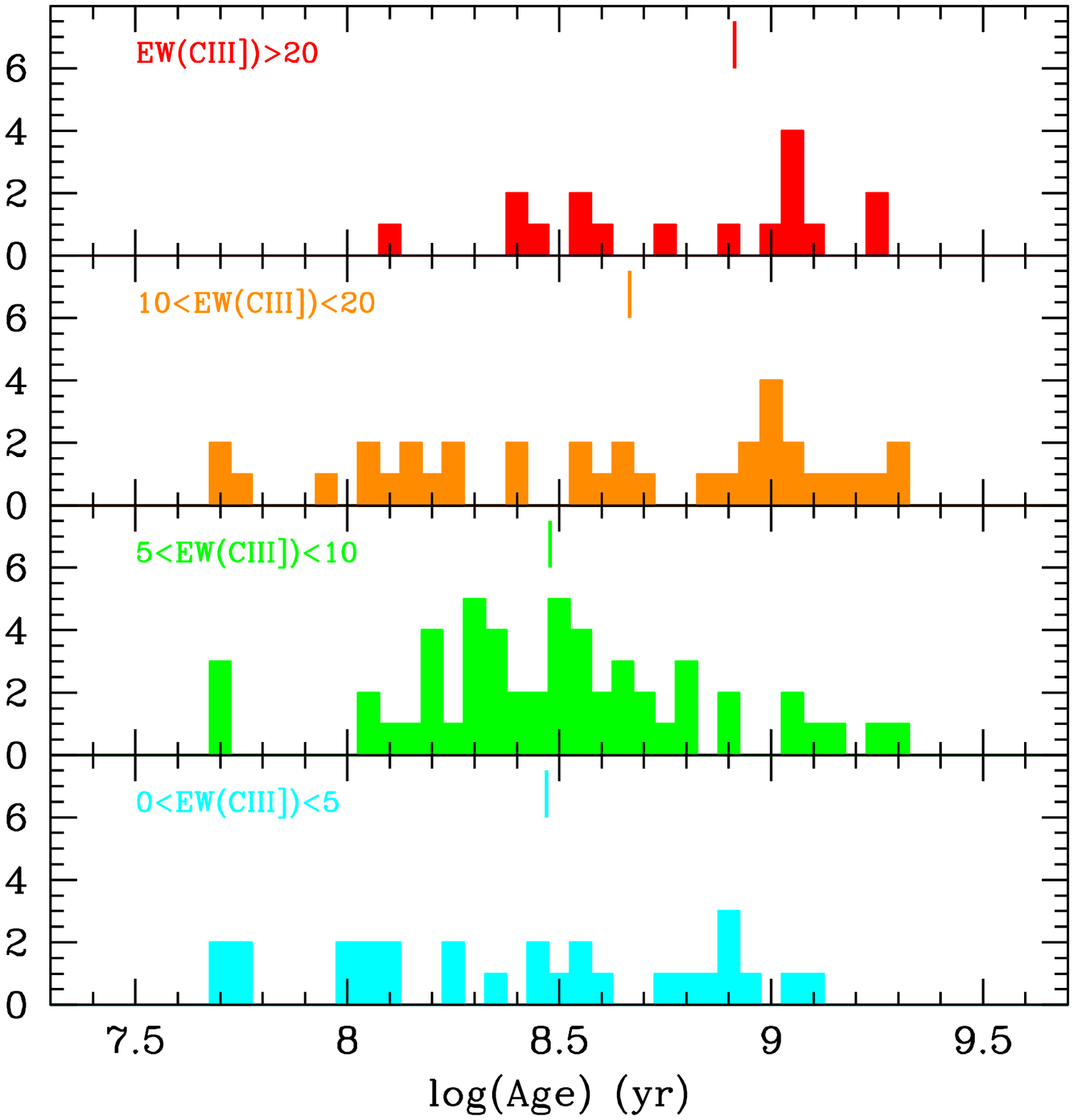}
      \caption{Age distribution of CIII] emitters with different EW(CIII]). The median age of each population is identified by the vertical lines color as for their respective sub-samples. The strength of the CIII] emission increases with the mean age.
}  
         \label{c3_age}
   \end{figure*}

We further investigate the properties of the CIII] emitters looking at the age distribution of the different categories, as presented in Fig.\ref{c3_age}, with ages computed  in \cite{Thomas2016} from the joint fitting of VUDS spectra and rest-UV and optical photometry. We use as an age definition the time since the onset of the last star formation event which dominate the star formation history, as modeled by a delayed exponential  \citep[see ][]{Thomas2016}. The median age of the galaxies is about the same for the weak and medium emitters, with 0.29$\pm$0.07 and 0.30$\pm$0.05 Gyr respectively, it increases  to  0.46$\pm$0.09 Gyr for the strong emitters, and reaches up to 0.82$\pm$0.13 Gyr for the very strong emitters. This relative age difference does not qualitatively change when using other age definitions like the mass-weighted age.

As pointed out in Sect.\ref{survey}  as AGN might not dominate the age sensitive regions of the SED,  the presence of AGN is not expected to significantly alter the stellar mass or SFR measurements obtained from SED fitting \citep[see e.g.][]{Bongiorno2012} and the results presented here are thus qualitatively expected to hold. However, more sophisticated SED modeling will be in order when higher signal to noise data will become available to separate AGN from stellar populations.


\section{Evidence for AGN quenching star-formation }
\label{agn_quench}

Several feedback-related quenching processes have been proposed, including AGN feedback, stellar wind feedback from supernovae and massive stars, or environmental quenching \citep[e.g.][]{Croton:2006,Hopkins2008,Gabor:2010,Gabor:2011}. Various pieces of evidence  presented in this paper converge to point out that the dominant process producing the observed quenching of star formation  in strong CIII] emitters is AGN related.

Starting from the perspective of analyzing the population of CIII] emitters, we have uncovered several important and related observed properties. We find a population of galaxies with high EW(CIII]), reaching in excess of 20\AA ~(Sect.\ref{c3_pop}). These strong emitters require a strong ionizing field beyond what stellar populations can produce, but rather at a level to what can be produced by AGN \citep{Nakajima:2017}. Emission line properties  indicate that a fraction of these galaxies must host an AGN as discussed in Sect.\ref{sect_agn}, and as extensively studied with line ratio diagnostics in \citet{Nakajima:2017}. \citet{Nakajima:2017} use line diagnostics defined from a large grid of photoionization models including stellar populations and AGN to find that about a third of the strong and very strong emitters in VUDS require the presence of an AGN to explain the high EW(CIII]) and the position of individual emitters in line ratio diagnostic diagrams.  The strongest emitters present strong outflows with velocity above 800 km/s beyond what stellar winds can produce, and rather comparable to that of AGN \citep{Talia2016}. In addition, we find that the strongest CIII] emitters appear below the main sequence, with SFR lower by about 0.4 dex for the stronger emitters compared to the weak emitters, and  mean ages almost three times older  than the SFG population (Sect.\ref{main_seq}). The analysis of stacked X-ray  imaging presented in Sect.\ref{sect_agn} further indicates a marginal detection (2$\sigma$) for the strongest CIII] emitters consistent with the presence of a low luminosity AGN with L$_X$(2-10keV)$<$3$\times10^{42}$ erg/s hosted by those galaxies. 

These observed properties all converge to favor AGN feedback as the main physical process capable of  quenching the star formation in the strong and very strong CIII] emitters. In an empirical scenario the strong ionizing field is  produced by AGN hosted in the galaxies observed as CIII] emitters. The jets and winds associated to the AGN would produce strong feedback, depositing energy in the ISM, gradually removing the gas supply and effectively quenching star formation, driving those galaxies below the MS. The large age difference of up to a factor of three observed between the CIII] emitters and main sequence galaxies may then indicate that the AGN feedback is taking at most a few hundred million years to operate. 
We therefore infer that part of the star-forming population is undergoing AGN feedback with the quenching of star formation,  at any given time 10.5 to 12.2 Gyr after the Big Bang. 

With one third of the strong and very strong CIII] emitters in the SFG sample which must host an AGN \citep{Nakajima:2017},  AGN feedback then concerns about one third of the 4\% of the SFG population that we identify with EW(CIII])$>$10\AA. If the duration of AGN activity at an energy level required to produce galaxy-wide quenching star-formation is less than the cosmic time covered by our observations, the fraction of the SFG population experiencing AGN feedback and quenching could be significantly higher. We estimate the quenching timescale assuming that SFGs subject to AGN feedback will then transform into the quiescent galaxies observed at $z\sim2$.  The volume density of quiescent galaxies with a stellar mass $M_{star}>10^{10}$M$_{\odot}$ at $z\sim2$ is reported to be $\sim 3\times10^{-3}$ gal/Mpc$^3$ \citep{Brammer2011,Ilbert:13,Barro2013}. From the analysis in this paper, one third of 4\% of SFGs above this mass translates into $3.3\times10^{-4}$ gal/Mpc$^3$.  For SFGs at a redshift $z\sim3$ to transform into the right number of quiescent galaxies over a cosmic time of 1.1 Gyr to $z\sim2$ then requires a quenching timescale $T_{quench}=(3.3\times10^{-4}/3\times10^{-3}) \times 1.1$ Gyr, or about 0.35Gyr, with a range $0.25-0.47$ Gyr given uncertainties on the AGN fraction. The effective timescale for star-formation quenching  from AGN feedback is currently highly uncertain from one hundred million years up to several hundred million years \citep[, see e.g.][]{Cattaneo:2009,Kaviraj:2011,Schawinski:2014}. Our results therefore seem to favor a long quenching timescale, as timescales much shorter than 0.35 Gyr would seem to overproduce quiescent galaxies by $z\sim2$. However, these numbers are indicative and will need to be refined with improved statistics on both the fraction of SFGs subject to AGN feedback, as well as on galaxies truly quiescent by $z\sim2$. Those numbers could also need to be adjusted if AGN feedback is only one of several ways to quench star formation for galaxies with $M_{star}>10^{10}$M$_{\odot}$.

Our results offer additional evidence for the co-evolution of AGN and star-formation at $z\sim2-4$. Star formation is observed to be present coevally with AGN activity \citep[e.g.][]{Fabian:2012,Cimatti:2013,Karman2014,Lemaux:14,Ishibashi:2016},  and this is what we are witnessing in the strong CIII] emitters population. 


\section{Discussion and summary}
\label{summary}

In this paper we present a statistical analysis of the properties of CIII]-$\lambda$1908 emitters with two aims: investigate the prevalence of CIII] emission in the star-forming galaxy population, and characterize the properties of the population of CIII] emitters using CIII] as a diagnostic tool, using measurements from the large sample of 3899 galaxies with $2<z<3.8$ in the VIMOS Ultra-Deep Survey (VUDS), .

The VUDS sample is selected from UV rest-frame luminosity ($i_{AB}\leq25$) and provides a large unbiased sample of UV-selected star-forming galaxies in a large volume minimizing cosmic variance \citep{LeFevre:15}, ideally suited to search for CIII] emitters, explore their properties and derive the frequency of appearance of this emission line. Spectral line measurements including lines from Ly$\beta$ up to CIII] is made possible from the deep VIMOS spectra. In addition, spectral model fitting performed simultaneously on spectra and multi-band photometry  \citep{Thomas2016} provides physical parameters including stellar mass, SFR, age, and dust extinction, which are used to characterize the population of CIII] emitters.

We split our sample in four categories of CIII] emitters spanning a range from weak emitters  0$<$EW(CIII])$_{rest}$$<$5\AA ~to very strong emitters with observed  EW(CIII]) up to $\simeq40$\AA ~for narrow line emission spectra. Individual spectra are stacked to produce average spectral properties for each category (Figures \ref{stacks_c3} and \ref{best}). We find that, as EW(CIII]) increases, emission lines requiring higher ionizing potential appear, with the strongest emitters showing NV-$\lambda$1240, SiII-$\lambda$1309, SiIV-$\lambda$1403, NIV]-$\lambda$1485, CIV-$\lambda$1549, HeII-$\lambda$1640, OIII]-$\lambda$1663, NIII]-$\lambda$1750, SiIII]-$\lambda$1888 (Fig. \ref{best}). 

We find a correlation between Ly$\alpha$ emission and CIII] emission (Figure \ref{lya_c3}), but with a large dispersion as we find strong CIII] emitters with Ly$\alpha$ in absorption, or strong Ly$\alpha$ emission and no or weak CIII]. We explore the properties of the CIII] population and find that EW(CIII]) weakly increases with near-UV luminosity L$_{NUV}$, and that the total size of galaxies gets smaller as EW(CIII]) increases (Fig.\ref{c3_prop}).

From the complete VUDS sample we compute the distribution of EW(CIII]) (Fig.\ref{dist_c3}) and find that about 24\% of the population shows some sign of CIII] emission. A first visual impression was already given in the Figure 10 of \cite{LeFevre:15}, as CIII] can be followed clearly from the 2D spectra image of  the whole VUDS sample. We quantify this statement, finding that 20\% of the population shows weak emission with $3<EW(CIII])<10$\AA, 4\% has $EW(CIII])>10$\AA, and 1.2\% presents very strong emission with $EW(CIII])>20$\AA. We find that LAE have a 42\% chance to show CIII] in emission with EW(CIII])$>$3\AA.  

While the presence of CIII] emission therefore appears rather frequent, we note that the vast majority of CIII] emitters are weak emitters making this line difficult to identify, except for the $\sim4$\% of the population with $EW(CIII])>10$\AA. One therefore cannot expect to rely on CIII] emission alone as a means to measure the spectroscopic redshifts of  galaxies in a complete sample of SFGs. Redshift measurements of UV-rest spectra without Ly$\alpha$ emission will still need to rely on a combination of the most prominent spectral features. Typically, besides CIII], lines like CIV, HeII, the continuum break at and below Ly$\alpha$, and hard-to-detect absorption features can jointly serve to eliminate degenerate redshift solutions.    

The fraction of strong CIII] emitters is  expected to increase  at higher redshifts towards the epoch of HI reionization (EoR) at $z>6$, only in the presence of a stronger ionizing field than observed in our sample at $2<z<3.8$, together with very low metallicity. Such conditions may exist in the lower luminosity galaxies in our sample (Sect.\ref{all_prop}) and such galaxies may be prevalent in the EoR. Some  analogs to those expected in the EoR have been discussed in  \citet{Amorin2017}. It is well possible that as the population of forming galaxies gets younger, with lower stellar mass, and lower metallicity, the fraction of objects with strong CIII] emission  in the EoR would be higher than in our sample at $z\sim3$.   This may be the case for the very first generation of stars in the first galaxies \citep{Nakajima:2017}.   Exactly how this fraction of strong emitters will increase at higher redshifts remains to be investigated.

The  main result from this paper is a coherent set of observational evidence pointing to AGN feedback acting on the population of strong CIII] emitters. Upon examination of spectral features, we identify four galaxies with broad emission lines, their quasar-like spectra clearly powered by strong AGN. We identify  7 objects with NV-$\lambda$1240 narrow emission line EW(NV)$>$10\AA ~recognized as a potential tracer of AGN activity.  The most striking observational results include the clear evidence for strong outflows up to $\sim1000$ km/s for the strongest CIII] emitters, and increasingly lower SFR in the SFR-M$_{star}$ plane as EW(CIII]) increases. Such strong outflows require energy levels beyond what stellar winds are capable to produce on a galaxy scale but are consistent with being powered by AGN \citep[e.g.][]{Harrison2012,Cimatti:2013,FS2014,Brusa2015}.  The distribution of CIII] emitters in the stellar mass vs. SFR plane around the main sequence of star-forming galaxies is investigated in Fig.\ref{ms_c3}. Our analysis shows that the strong and very strong emitters are on average below the MS indicating that the star formation is quenched in these galaxies, with the SFR decreasing when EW(CIII]) increases. From X-ray data stacking, we find a marginal 2$\sigma$ detection consistent with low luminosity AGN  with L$_X$(2-10 keV) $\sim$42.9 erg/s for the strong and very strong CIII] emitters, as observed on the faint end of the AGN luminosity function at these redshifts \citep{Georgakakis:2015}.

The physical origin of CIII] emission in relation to the prevalent ionizing field is investigated in \citet{Nakajima:2017}. From the data presented here, \citet{Nakajima:2017} derive various line ratio diagnostics on the basis of a large range of photoionizing models combining CIII]-$\lambda$1908 with other emission lines like CIV-$\lambda$1549, HeII-$\lambda$1640, OIII]-$\lambda$1664. It is found that the strong and very strong CIII] emitters must host an AGN in order to produce the large EW(CIII]) and observed line ratios.
The spectral analysis in \citet{Nakajima:2017} therefore reinforces our proposed scenario with strong AGN-driven quenching of star-formation. Without AGN, it seems unlikely to find conditions with the high star-formation activity necessary to produce strong CIII] and the observed line ratios, while at the same time producing strong outflows and driving galaxies off the MS. We rather conclude that it is more realistic that AGN and star formation activity act together to produce the observed properties of CIII] emitters. 

The signature of AGN have been previously reported in SFGs at and beyond the peak in SFRD \citep[e.g.][]{Cimatti:2013,genzel2014,Talia2012,Karman2014,Talia2016}. The regulation of mass growth from the quenching of star-formation has been proposed to support an evolutionary path gradually transforming  SFGs at the highest redshifts into quiescent galaxies observed at $z\sim1-2$ \citep{Barro2013}, with AGN feedback  identified from numerical simulations as the prime contender for the quenching of star formation \citep{Hopkins2006,Hopkins2008}.  The results presented in this paper bring further evidence of the presence of AGN in SFGs at a key epoch of mass build-up in galaxies

The properties of CIII] emitters are consistent with an empirical scenario in which CIII] emitting galaxies with increasing EW host an increasingly larger fraction of AGN producing the strong ionizing field needed to ionize CIII] to the observed levels. Those AGN  drive the strong observed outflows, and in the end quench star-formation in the host galaxies to produce the observed location of CIII] emitters in the SFR-M$_{star}$ plane.  The age of the stellar populations of the stronger emitters is found to be about three times that of the weak emitters, at $\sim$0.8 vs. 0.3 Gyr, respectively. This may be an indication that the strongest emitters were the first to experience AGN feedback and associated quenching. We find that quenching timescales of $\sim$$0.25-0.5\times10^8$ years are required to transform SFGs with M$_{star}>10^{10}$M$_{\odot}$ into the volume density of quiescent galaxies observed at $z\sim2$ \citep{Brammer2011,Ilbert:13,Barro2013}. 

We conclude that the co-evolution of stellar populations and AGN is a key ingredient to the star formation evolution and mass build-up in galaxies at these early times. Future experiments assembling  larger datasets will be necessary to  investigate whether the AGN feedback and quenching identified in the redshift range of this paper and beyond is mass-dependent.


The results presented here point to the need for large comprehensive spectroscopic surveys with the power to identify galaxy populations which may appear as rare but provide important clues on the first phases of galaxy assembly. Our study paves the way to a systematic use of spectral line properties and diagnostics from the UV to optical rest-frame for future investigations using next generation facilities like the James Webb Space Telescope and ELTs.

\normalsize


\begin{acknowledgements}
This work is supported by funding from the European Research Council Advanced Grant ERC--2010--AdG--268107--EARLY and by INAF Grants PRIN 2010, PRIN 2012 and PICS 2013. 
AC, OC, MT and VS acknowledge the grant MIUR PRIN 2015.  
This work is based on data products made available at the CESAM data center, Laboratoire d'Astrophysique de Marseille, France. 
\end{acknowledgements}


\normalsize

\bibliographystyle{aa} 
\bibliography{vuds_c3} 


\newpage
\tiny

\begin{landscape}

\begin{table}
\caption{Equivalent width of emission lines in stacked CIII] spectra}
\label{stacks_prop}    
\begin{tabular}{|c|cc|cc|cc|cc|}       
\hline             
                                & \multicolumn{8}{c|}{Sample} \\ \hline  
Line-$\lambda$  & \multicolumn{2}{c|}{$EW(CIII]) \geq 20$\AA}  &  \multicolumn{2}{c|}{$10 \leq EW(CIII])<20$\AA}   &  \multicolumn{2}{c|}{$5 \leq EW(CIII])<10$} &  \multicolumn{2}{c|}{AGN type II}      \\ 
                               &    EW (\AA)   &   Flux (relative to CIII)                 &    EW (\AA) &   Flux (relative to CIII)                             &    EW (\AA) &   Flux (relative to CIII)                    &    EW (\AA)   & Flux (relative to CIII)   \\  \hline
Ly$\alpha$       &   $71.7_{-5}^{+4}$            &   $4.6\pm0.07$              &   $58.8_{-3.5}^{+7.0}$        &  $6.4\pm0.04$   &  $30.3_{-1.5}^{+2.3}$     & $5.2\pm0.2$ & $98.8_{-3}^{+3}$      &  $6.25\pm0.06$    \\
NV-$\lambda$1240\AA       &    $0.9_{-0.3}^{+0.3}$      &  $0.05\pm0.02$               &   $0.3_{-0.15}^{+0.15}$     &   $0.02\pm0.01$ &  $-$   				& $-$ & $19.1_{-2}^{+2}$      &  $1.31\pm0.1$    \\
SiII-$\lambda$1309\AA       &    $0.6_{-0.2}^{+0.2}$      &  $0.07\pm0.02$              &  $0.2_{-0.15}^{+0.15}$       &   $0.04\pm0.02$  &  $-$                                   & $-$  & $1.7_{-0.4}^{+0.4}$ &  $0.12\pm0.04$     \\
SiIV-$\lambda$1403\AA      &    $1.4_{-0.2}^{+0.2}$     &   $0.13\pm0.03$              &  $0.1_{-0.15}^{+0.15}$      &   $0.02\pm0.02$  &  $-$                                   & $-$  & $1.9_{-0.5}^{+0.4}$ &  $0.14\pm0.04$     \\
NIV-$\lambda$1485\AA      &    $2.2_{-0.3}^{+0.3}$      &  $0.19\pm0.02$              &   $0.6_{-0.1}^{+0.1}$         &   $0.08\pm0.02$   &  $-$                                   & $-$  & $2.5_{-0.5}^{+0.5}$  &  $0.19\pm0.05$     \\
CIV-$\lambda$1549\AA      &    $4.4_{-0.5}^{+0.6} $    &  $0.31\pm0.03$               &  $2.4_{-0.2}^{+0.5}$          &  $0.36\pm0.02$    &  $0.1_{-0.3}^{+0.2}$     & $0.01\pm0.1$ & $38.8_{-2.1}^{+2.0}$ &  $3.00\pm0.08$      \\
HeII-$\lambda$1640\AA     &     $4.3_{-0.4}^{+0.7}$    &  $0.27\pm0.03$                &  $3.2_{-0.3}^{+0.5}$         &    $0.49\pm0.03$ &   $1.7_{-0.2}^{+0.2}$     & $0.3\pm0.06$ & $11.7_{-0.5}^{+1.0}$ & $0.92\pm0.06$      \\
OIII]-$\lambda$1664\AA    &      $5.5_{-0.5}^{+0.6}$    &  $0.27\pm0.03$                &  $1.4_{-0.3}^{+0.4}$         &    $0.14\pm0.02$  &  $0.7_{-0.2}^{+0.2}$     & $-$ & $1.7_{-0.5}^{+0.6}$   & $0.13\pm0.05$     \\
NIII-$\lambda$1750\AA     &      $1.1_{-0.2}^{+0.2}$     &  $0.07\pm0.03$               &   $0.4_{-0.2}^{+0.2}$        &    $0.04\pm 0.02$  &  $-$                                   & $-$ & $1.5_{-0.6}^{+0.7}$  &  $0.13\pm0.06$     \\
SiIII-$\lambda$1888\AA     &     $5.0_{-0.3}^{+0.4}$      &  $0.18\pm0.12$               &  $4.0_{-0.5}^{+0.4}$         &    $0.31\pm0.10$  &  $-$                                   & $-$ & $2.5_{-0.8}^{+0.9}$   &   $0.19\pm0.02$     \\
CIII]-$\lambda$1908\AA    &      $23.5_{-1.8}^{+1.6}$    &  $1\pm0.05$                    &  $11.1_{-1.1}^{+1.2}$       &   $1\pm0.04$       &    $7.1_{-1.1}^{+1.3}$      & $1\pm0.15$ & $13.2_{-1.0}^{+1.2}$ & $1.0\pm0.09$         \\ \hline
UV slope $\beta$      &      \multicolumn{2}{c|}{$-1.76\pm0.08$}     &  \multicolumn{2}{c|}{$-1.61\pm0.03$}  &  \multicolumn{2}{c|}{$-1.35\pm0.02$} &  \multicolumn{2}{c|}{$0.09\pm0.02$}      \\ \hline 
$\Delta$V$_{outflow}$ km/s      &      \multicolumn{2}{c|}{$1014\pm205$}     &  \multicolumn{2}{c|}{$445\pm110$}  &  \multicolumn{2}{c|}{$137\pm61$} &  \multicolumn{2}{c|}{--}      \\ \hline 
\end{tabular}
\end{table}

\normalsize

\begin{table}
\caption{Equivalent width of emission lines in stacked CIII] spectra}
\label{stacks_sfg}    
\begin{tabular}{|c|cc|cc|}       
\hline             
                                & \multicolumn{4}{c|}{Sample} \\ \hline  
Line-$\lambda$  & \multicolumn{2}{c|}{SFGs $2<z<3$ }  &  \multicolumn{2}{c|}{SFGs $3<z<4$}      \\ 
                             &    EW (\AA)   & Flux (relative to CIII)                     &    EW (\AA)   & Flux (relative to CIII)  \\  \hline
Ly$\alpha$                          & $7.9_{-1.3}^{+1.3}$ & $3.5\pm0.1$ &   $14.8_{-1.2}^{+1.5}$  & $4.5\pm0.12$   \\
CIV-$\lambda$1549\AA      & $2.78_{-0.06}^{+0.11}$  & $-2.1\pm0.05$ &  $2.84_{-0.05}^{+0.15}$  & $-1.2\pm0.05$   \\
CIII]-$\lambda$1908\AA    & $2.0_{-0.2}^{+0.2}$ & $1\pm0.05$  &  $2.24_{-0.2}^{+0.3}$      &  $1\pm0.05$  \\ \hline
UV slope $\beta$              &  \multicolumn{2}{c|}{$-0.94\pm0.02$} &  \multicolumn{2}{c|}{$-0.92\pm0.02$} \\ \hline 
$\Delta$V$_{outflow}$ km/s      & \multicolumn{2}{c|}{$65\pm41$} & \multicolumn{2}{c|}{$98\pm40$}   \\ \hline 
\end{tabular}
\end{table}

\end{landscape}

\end{document}